%% file: systematics.tex
\documentclass[reprint, prd, superscriptaddress, tightenlines, longbibliography, nofootinbib, eqsecnum, amsfonts, amsmath, floatfix, notitlepage]{revtex4-1}
\pdfoutput=1

\usepackage[utf8]{inputenc}
\usepackage{mathrsfs}
\usepackage{euscript}
\usepackage{graphics}
\usepackage{graphicx}
\usepackage{amsmath}
\usepackage{amssymb}
\usepackage{gensymb}
\usepackage{bm}
\usepackage[usenames,dvipsnames,svgnames,table]{xcolor}
\definecolor{linkcolor}{rgb}{0.6,0,0}
\definecolor{citecolor}{rgb}{0,0,0.75}
\definecolor{urlcolor}{rgb}{0.12,0.46,0.7}
\usepackage[breaklinks, colorlinks, urlcolor=urlcolor, linkcolor=linkcolor,citecolor=citecolor,pdfencoding=auto]{hyperref}
\usepackage{xspace}
\usepackage{wasysym}
\usepackage{times}
\usepackage{appendix}
\usepackage{comment}
\usepackage{lipsum}
\usepackage[nolist,nohyperlinks]{acronym}
\usepackage{float}
\usepackage{simplewick}
\usepackage{natbib, ifthen}
\usepackage{longtable}
\usepackage{mhchem} 
\usepackage[caption=false]{subfig}
\usepackage{multirow}
\usepackage{nameref}
\usepackage{bbold}
\usepackage{soul}

\input macros.tex

\newcommand{\beq}{\begin{equation}}
\newcommand{\enq}{\end{equation}}
\newcommand{\beqa}{\begin{eqnarray}}
\newcommand{\enqa}{\end{eqnarray}}
\newcommand{\beit}{\begin{itemize}}
\newcommand{\enit}{\end{itemize}}
\newcommand{\bem}{\begin{pmatrix}}
\renewcommand{\enm}{\end{pmatrix}}

\newcommand{\vecL}{\mathbf L}

\newcommand{\vecx}{\mathbf{x} }
\newcommand{\vecy}{\mathbf{y} }

\newcommand{\obs}{\textrm{obs}}

\newcommand{\eff}{\textrm{eff}}

\newcommand{\fpatch}{f_{A,L}}
\renewcommand{\max}{\mathrm{max}}
\renewcommand{\min}{\mathrm{min}}
\newcommand{\maj}{\mathrm{maj}}

\renewcommand{\bem}{\begin{bmatrix}}
\renewcommand{\enm}{\end{bmatrix}}

\newcommand{\nhat}{{\bf\hat{n}}}

\newcommand{\Noise}{ {\boldsymbol N}}

\newcommand{\vecell}{ {\boldsymbol \ell}}

\newcommand{\FWHM}{{\rm FWHM}}

\newcommand{\sfcmb}{\texttt{s4cmb}\xspace}
\newcommand{\namaster}{\texttt{NaMaster}\xspace}

\newcommand{\dat}{\rm dat}

\newcommand{\MC}{\text{MC}}

\newcommand{\RDN}{\ce{^{\text{(RD)}}$N$}}
\newcommand{\MCN}{\ce{^{\text{(MC)}}N}}
\newcommand{\pC}{\tilde{C}}
\newcommand{\fid}{\rm fid}

\newcommand{\mb}{\bm{b}}

\newcommand{\filt}{\rm filt}

\newcommand{\corr}{\rm corr}

{ 
\newcommand{\MF}{\rm MF}

\newcommand{\R}{\mathcal{R}}

\newcommand{\leak}{\rm leak}
\renewcommand{\t}{\rm top}
\renewcommand{\b}{\rm bottom}
\newcommand{\N}{\mathcal{N}}
\renewcommand{\U}{\mathcal{U}}

\newcommand{\fmux}{{\rm fMUX}}
\newcommand{\mmux}{\mu{\rm MUX}}
\newcommand{\vex}{\boldsymbol{X}}

\newcommand{\hats}{\hat{\vs}}
\renewcommand{\mod}{\; {\rm mod} \;}
\newcommand{\beam}{\rm beam}
\newcommand{\mux}{\rm MUX}
\newcommand{\CS}{{\rm cs}}
\newcommand{\beamed}{{\rm b}}
\newcommand{\lb}{L_{\rm bin}}
\newcommand{\syst}{{\rm syst}}
\newcommand{\alphacovec}[3]{{\boldsymbol \alpha_{\bf #1}^{\bf #2#3}}}
\newcommand{\alphaco}[1]{{\alpha_{#1}}}
\newcommand{\Rvec}{{\bf R}}
\newcommand{\sigmavec}{{\boldsymbol\sigma}}
\newcommand{\cvec}{\boldsymbol{C}}
\newcommand{\theory}{{\rm theory}}

\let\oldsqrt\sqrt
\def\sqrt{\mathpalette\DHLhksqrt}
\def\DHLhksqrt#1#2{%
\setbox0=\hbox{$#1\oldsqrt{#2\,}$}\dimen0=\ht0
\advance\dimen0-0.2\ht0
\setbox2=\hbox{\vrule height\ht0 depth -\dimen0}%
{\box0\lower0.4pt\box2}}

\usepackage[dvipsnames]{xcolor}

\graphicspath{{./Figures/}}
\tolerance=1
\emergencystretch=\maxdimen
\hyphenpenalty=10000
\hbadness=10000

\usepackage{tikz}
\usepackage{collcell}
\usepackage{etoolbox}

\newtoggle{inTableHeader}
\toggletrue{inTableHeader}
\newcommand*{\StartTableHeader}{\global\toggletrue{inTableHeader}}%
\newcommand*{\EndTableHeader}{\global\togglefalse{inTableHeader}}%

\let\OldTabular\tabular%
\let\OldEndTabular\endtabular%
\renewenvironment{tabular}{\StartTableHeader\OldTabular}{\OldEndTabular\StartTableHeader}%

\newcommand*{\brightnessthree}{White!30}%

\newcommand{\ApplyGradient}[1]{%
	\iftoggle{inTableHeader}{#1}{
		\pgfmathsetmacro{\PercentColor}{max(min(100.0*(#1),100.0),0.00)} %
			\hspace{-0.33em}\vspace{-0.03em}\colorbox{\brightnessthree!red!\PercentColor!}{#1}
}}

\setlength{\fboxsep}{3.5mm} 
\setlength{\tabcolsep}{-0.8pt}

\interfootnotelinepenalty=10000

\begin{document}
\newcommand{\Sussex}{Department of Physics \& Astronomy, University of Sussex, Brighton BN1 9QH, UK}
\newcommand{\Paris}{Universit{\'e} Paris-Saclay, CNRS/IN2P3, IJCLab, Orsay, France}
\newcommand{\Cardiff}{School of Physics and Astronomy, Cardiff University, The Parade, Cardiff, CF24 3AA, UK}
\newcommand{\CCA}{Center for Computational Astrophysics, Flatiron Institute, 162 5th Avenue, 10010, New York, NY, USA}
\author{Mark Mirmelstein}
\email{M.Mirmelstein@sussex.ac.uk}
\affiliation{\Sussex}
\author{Giulio Fabbian}
\affiliation{\CCA}
\affiliation{\Cardiff}
\affiliation{\Sussex}
\author{Antony Lewis}
\affiliation{\Sussex}
\author{Julien Peloton}
\affiliation{\Paris}
\affiliation{\Sussex}


\title{Instrumental systematics biases in CMB lensing reconstruction: a simulation-based assessment}

\date{\today}
\begin{abstract}
Weak gravitational lensing of the cosmic microwave background (CMB) is an important cosmological tool that allows us to learn about the structure, composition and evolution of the Universe. Upcoming CMB experiments, such as the Simons Observatory (SO), will provide high-resolution and low-noise CMB measurements. We consider the impact of instrumental systematics on the corresponding high-precision lensing reconstruction power spectrum measurements. We simulate CMB temperature and polarization maps for an SO-like instrument and potential scanning strategy, and explore systematics relating to beam asymmetries and offsets, boresight pointing, polarization angle, gain drifts, gain calibration and electric crosstalk. Our analysis shows that the majority of the biases induced by the systematics we modeled are below a detection level of $\sim 0.6\sigma$. We discuss potential mitigation techniques to further reduce the impact of the more significant systematics, and pave the way for future lensing-related systematics analyses.
\end{abstract}

\maketitle

\section{Introduction}
\label{sec:intro}

One of the main scientific objectives of upcoming cosmic microwave background (CMB) experiments is to measure the gravitational lensing of the CMB photons over a substantial sky area with the highest precision to date. This will enable us to better constrain dark energy models and inflation, provide more information on neutrino masses, and learn more about the large-scale structure of the Universe up to high redshift. To achieve this from upcoming observations, it is crucial to understand how instrumental systematics could bias the lensing potential reconstruction. This challenge will be more important for future CMB experiments such as the Simons Observatory (SO)~\cite{Ade:2018sbj} and CMB-S4~\cite{Abazajian:2016yjj}, as small systematics become more significant with higher resolution and lower noise levels.

The ability to effectively reconstruct the lensing potential (see~\cite{Lewis:2006fu} for a review) from upcoming ground-based CMB experiments could be limited by various instrumental systematics. For example, systematics could induce lensing-like features in the CMB maps or act to effectively increase the reconstruction noise. Previous work in the literature has characterized the influence and potential significance of several systematics on lensing reconstruction~\cite{Hu:2002vu,Miller:2008zi,Su:2009tp}. However, these treatments have mostly used analytic approximations and idealized scanning strategies, rather than employing realistic instrument, scans, and modeling of systematics based on levels observed in real data. Several experiments have used a simulation-based approach to characterize residual systematic uncertainties, but focused on the CMB power spectra~\cite{Ade:2017uvt,Akrami:2018vks,Pagano:2019tci} or to guide the design of future instruments~\cite[e.g.][]{Crowley:2018eib}. The POLARBEAR collaboration has recently used a simulation-based approach to propagate residual systematics uncertainties in their latest lensing reconstruction measurements~\cite{pblensing2020}.

In this work, we adopt a similar end-to-end simulation approach to propagate the most common instrumental effects related to beam, calibration, pointing and readout electronics through to a lensing reconstruction analysis for a next-generation SO-like instrument. We assume a realistic amplitude for the modeled systematics, as observed in the current generation of experiments, or as expected for the next generation instruments given their design specifications. Our baseline reconstruction pipeline performs a semi-optimal treatment of noise inhomogeneities induced by the scanning strategies of ground-based experiments~\cite{Mirmelstein:2019sxi}, however it does not automatically mitigate possible residual systematic biases. Using systematics-free Monte Carlo (MC) simulations to obtain the noise-debiasing terms for the CMB lensing power spectrum could lead to biases on real data. It is therefore crucial to understand the detailed behavior of systematics-induced biases and their potential significance for a more accurate lensing reconstruction, and to design mitigation strategies when required (for example, by including an accurate model of the most important effects in the corresponding MC simulations used in the lensing analysis).
In this work, we focus on the CMB lensing power spectrum reconstruction, and do not consider the impact on other important analyses such as the delensing of CMB polarization. A more detailed study may be required for the delensing analysis, as small systematics-induced map-level effects could become relatively more important.

This paper is structured as follows. We begin in Sec.~\ref{sec:instrument_systematics_simulations} by describing how we model the instrumental systematics that we consider, and how these systematics affect the time-stream data simulations. Sec.~\ref{sec:lensing_analysis} gives a short overview of the lensing reconstruction pipeline that we use to analyze our simulations. The systematics-induced biases and their significance on the CMB power spectra and the reconstructed lensing power spectrum are shown in Sec.~\ref{sec:systematics_biases}. We discuss possible mitigation strategies in Sec.~\ref{sec:mitigation_techniques}, and summarize our findings and future prospects in Sec.~\ref{sec:conclusions_and_future_prospects}. Throughout this paper we assume a Gaussian unlensed CMB model corresponding to a fiducial~$\Lambda$CDM model with Planck-estimated parameters~\cite{Ade:2015xua}, and inhomogeneous but pixel-uncorrelated instrumental noise (e.g. neglecting $1/f$ noise from the atmosphere or from instrument electronics). We also do not attempt to model systematics that couple to foregrounds, and consider an experiment which is insensitive to the CMB temperature monopole and dipole. For the scanning strategy considered here, the latter are largely removed by the filtering usually employed on real data to handle slowly varying correlated ($1/f$) noise induced by the atmosphere.

\section{Instrumental systematics simulations}
\label{sec:instrument_systematics_simulations}

Due to the time dependency of the data acquisition chain of CMB experiments, the most accurate and natural way to include the effects of instrumental systematics is to inject them at the raw time-ordered data (TOD) level. For this purpose, and to construct sky maps from the simulated TOD, we use the public Python package \sfcmb\footnote{Available at: \href{https://github.com/JulienPeloton/s4cmb/}{https://github.com/JulienPeloton/s4cmb/}.}~\cite{Fabbian:2021hlw}. This software, which is derived from the POLARBEAR data analysis systematics pipeline, has been used to perform a preliminary systematics study for SO~\cite{Crowley:2018eib,Gallardo:2018rix,Salatino:2018voz}, and to explore the effects of systematics on $B$-mode measurements on real data~\cite{Ade:2017uvt,pb2014}. Injecting systematics directly into the simulated detector-by-detector TOD allows us to explore a wider set of systematics in a more realistic way than other possible treatments of systematics (such as effective induced map-domain systematics), and includes their variation across the focal plane of the instrument.

We start by converting noise-free and beam-free CMB temperature and polarization realization maps, $\vs=\{T,Q,U\}$, to TOD based on instrument specifications. The instrument (white) noise $\vn$, instrument beam $b$, and systematics, are then injected into the TOD, which is then converted to temperature and polarization maps following a scanning strategy's pointing model.
We can write the generated data time stream $d_t$ for a specific time sample $t$ as
\begin{equation}
d_t = T_t + Q_t\cos(2\psi_t) + U_t\sin(2\psi_t)+n_t,
\label{eq:tod}
\end{equation}
where $\psi$ is the polarization angle of the detector with respect to the sky coordinates, $T_t$, $Q_t$ and $U_t$ are the $T, Q, U$ Stokes parameters of the CMB observed in the sky direction where the telescope is pointing at given time $t$, and $n$ is the instrument noise. At this point we define the CMB signals to already be affected by the instrument beam and systematic effects. The instrument noise, which is not affected by the beam, may also be affected by some systematic effects such as gain variations. Throughout, we hereafter drop the $t$ subscript for convenience. The way in which the systematics we model affect the TOD is shown individually in the following subsections.

The generated TOD with systematics is then converted into three temperature and polarization flat-sky maps using a binned map-making process; rewriting Eq.~\eqref{eq:tod} in vector notation,
\begin{equation}
\vd = \mA\vs+\vn,
\label{eq:map_making}
\end{equation}
where $\mA$ is the pointing matrix of the scanning strategy, the reconstructed sky maps $\hats$ are the generalized least square solution of Eq.~\eqref{eq:map_making}~\cite{PhysRevD.56.4514,Stompor:2002jy},
\begin{equation}
\hats = \left({\mA}^{\top}\mN^{-1} \mA\right)^{-1} {\mA}^{\top}\mN^{-1}\vd,
\label{eq:map_making_solution}
\end{equation}
where $\mN$ is the time-domain instrument noise correlation matrix that we assume is diagonal and proportional to the noise variance of the TOD (and the same for all the detectors). In the following we will use a pair-differencing approach, where we map independently the half sum and half difference of the TOD from a pair of detectors within a focal plane pixel that observe the sky with orthogonal polarization angles~\cite{Jones:2006ac}. This is a commonly-used strategy to isolate the polarized and unpolarized components of the signal while minimizing the mixing between the two. Mixing between intensity and polarization is particularly dangerous for polarization measurements from the ground, for which any leakage of the unpolarized signal is dominated by the strong atmospheric emission. We do not use any filtering during the map-making process to avoid the need to correct the reconstructed lensing potential power spectrum by additional MC corrections due to filter-induced biases. While some filtering procedures~\cite{Stompor:2002jy,Poletti:2016xhi} or other map-making-stage modifications~\cite{Adam:2015vua,Wallis:2015ypa,Delouis:2019bub} may mitigate some systematic effects, in this work we only demonstrate to leading order the potential lensing biases which may result from systematics alone.

We use the process described above to obtain three groups of simulations:
\begin{enumerate}
\item \textit{MC simulations}: systematics-free simulations which are obtained using our default instrument specifications and scanning strategy. These make up different simulation sets used for calculating different debiasing terms for the lensing reconstruction analysis. In total, we use 576 MC simulations. Their allocation to the different debiasing terms is described in Sec.~\ref{sec:lensing_analysis}.
\item \textit{Systematics-free ``data'' simulations}: 10 simulations similar to the MC simulations, but using a specific set of 10 CMB + noise realizations. These simulations are used for a systematics-free lensing reconstruction analysis for comparison.
\item \textit{``Data'' simulations}: same as group 2, but with the effect of systematics. Each considered systematic has its own set of 10 ``data'' simulations from which we reconstruct the lensing potential. The averaged reconstructed lensing power spectra of this set are compared to the same power spectra obtained from the systematics-free ``data'' set.
\end{enumerate}

\begin{figure}[!h]
\centering
\includegraphics[width=\columnwidth]{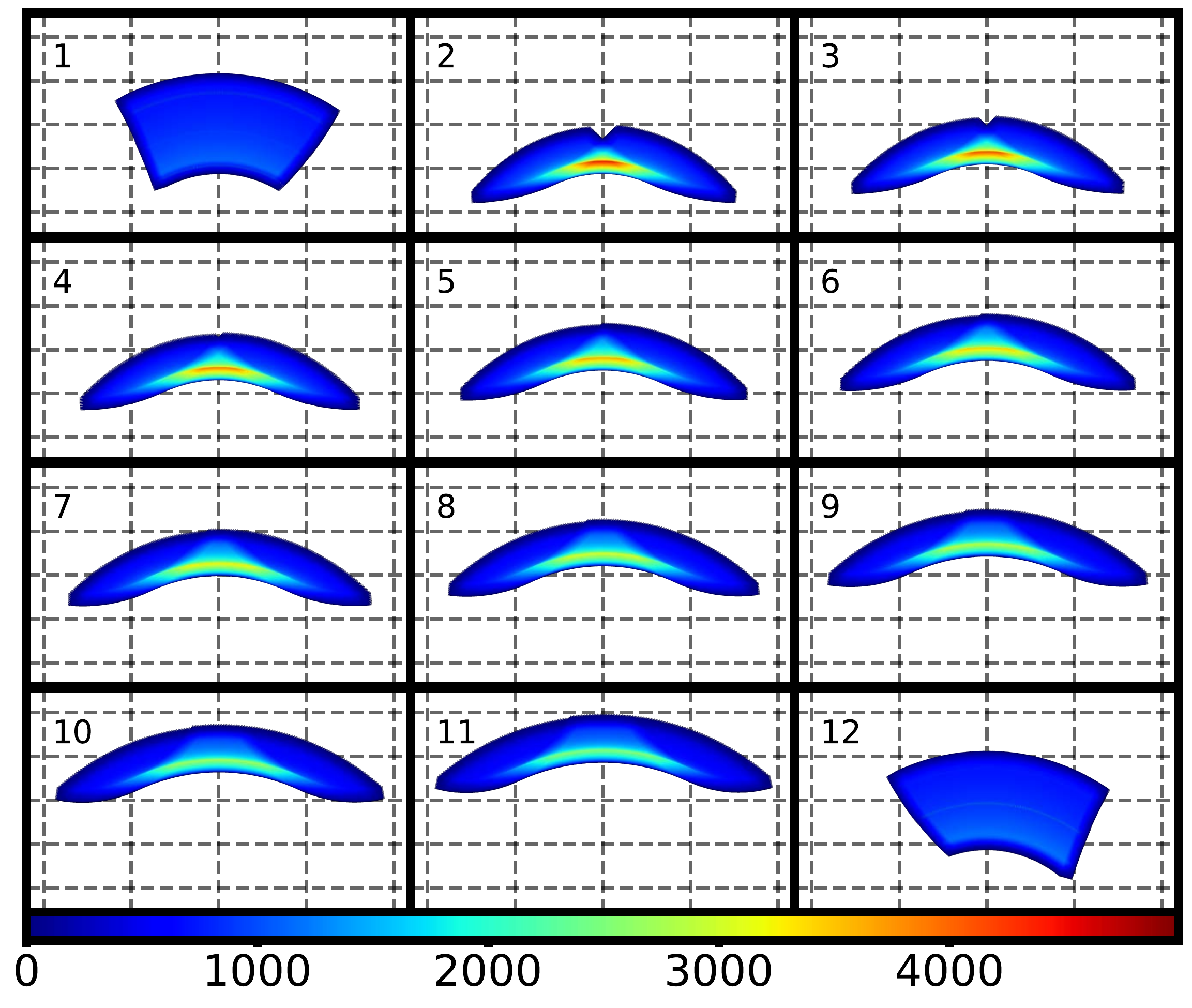}
\caption{Sky coverage of each constant elevation scan (CES) of our scanning strategy. Each of scans 2-11 simulates $\sim4$ observation hours while scans 1 and 12 simulate $\sim5$ hours. In each CES, all detectors in the focal plane operate at the same time. The color map shows the number of observations per pixel in the scanned regions. Blue areas are observed less, and red areas are observed more times. The sub-panels of this figure cover the same area of Fig.~\ref{fig:weights}, where we show the full composition of the scans.}
\label{fig:CES}
\end{figure}

To simulate realistic observations, we use an existing scanning strategy in \sfcmb, the ``deep patch'' scan, for all simulations. This simulates observations covering $\sim$5\% of the sky, which is consistent with the plan of the deepest CMB observations of SO and CMB-S4~\cite{Stevens:2018biw,s4-rforecast}. The specific scanning strategy we adopted throughout this work is composed of 12 individual constant elevation scans (CESs) having a unique scanning pattern, as shown in Fig.~\ref{fig:CES}. Future surveys dedicated to CMB lensing science will typically cover a much larger sky area ($\sim50\%$ of the sky). However, our specific choice of scanning strategy is a good compromise that allows us to perform rapid simulations relatively inexpensively numerically. As we will discuss later on, the amount of cross-linking of the scans is a crucial factor affecting the impact of several instrumental systematics.

\begin{figure}[!h]
\centering
\includegraphics[width=\columnwidth]{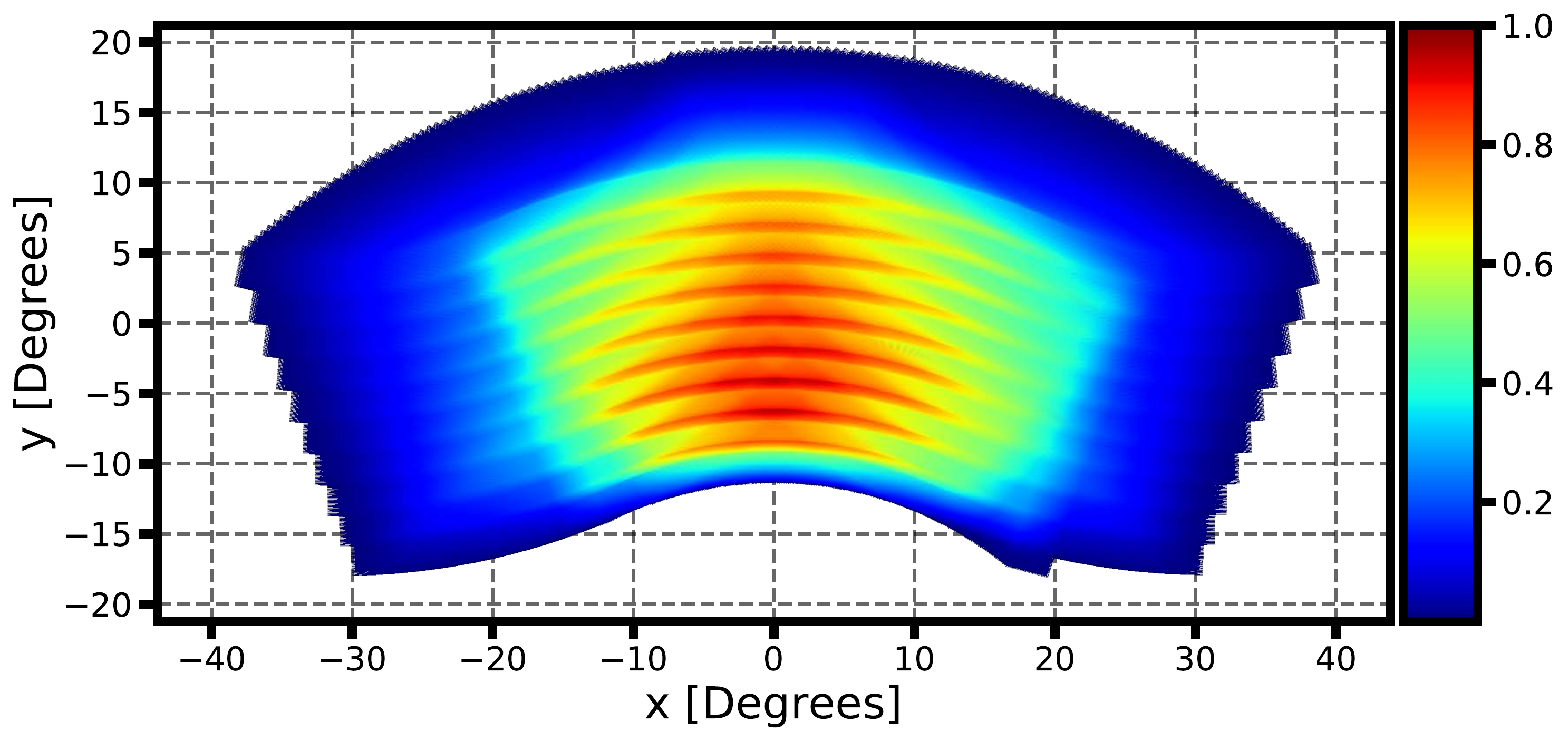}
\caption{The full normalized hit count map composed of the different CESs from Fig.~\ref{fig:CES}. This hits map is the baseline for constructing the anisotropic noise covariance map which is used in the lensing reconstruction filtering processes. Blue areas are observed less, and red areas are observed more times. The resolution of the maps is 1.7 arcminutes and the total sky area observed is $\sim$5\% of the full sky.}
\label{fig:weights}
\end{figure}

The normalized hit count map of the full 12-day scan is shown in Fig.~\ref{fig:weights}. This ``weights'' map is used throughout the lensing analysis as a baseline for the anisotropic noise covariance map. A hit, or \textit{sample}, is acquired every 1/15 seconds with a telescope's constant azimuth speed of $0.4\degree/sec$ at an elevation of 5,200 meters for an observatory located in the Atacama plateau (which is the SO location). Since we simulate only 12 days of observations, which is only a fraction of a CMB experiment's full multiyear run, the effects of systematics which are expected to average out with time will be larger than in reality. Our bias estimates from these systematics should therefore be closer to an upper bound for what a similar experiment might observe in reality.

While our chosen scanning strategy is commonly employed by ground-based CMB experiments, other scans may be more optimal for mitigating systematics~\cite{Thomas:2019hak}. We focus on the scan defined above to characterize any lensing biases, so we can understand in a baseline configuration which systematics may be important for upcoming CMB experiments, and hence require more detailed study. Using a simple scan also avoids underestimating biases due to the choice of a specific more-complex scanning pattern that may not actually be implemented by future experiments.

\begin{figure}[!h]
\centering
\includegraphics[width=0.8\columnwidth]{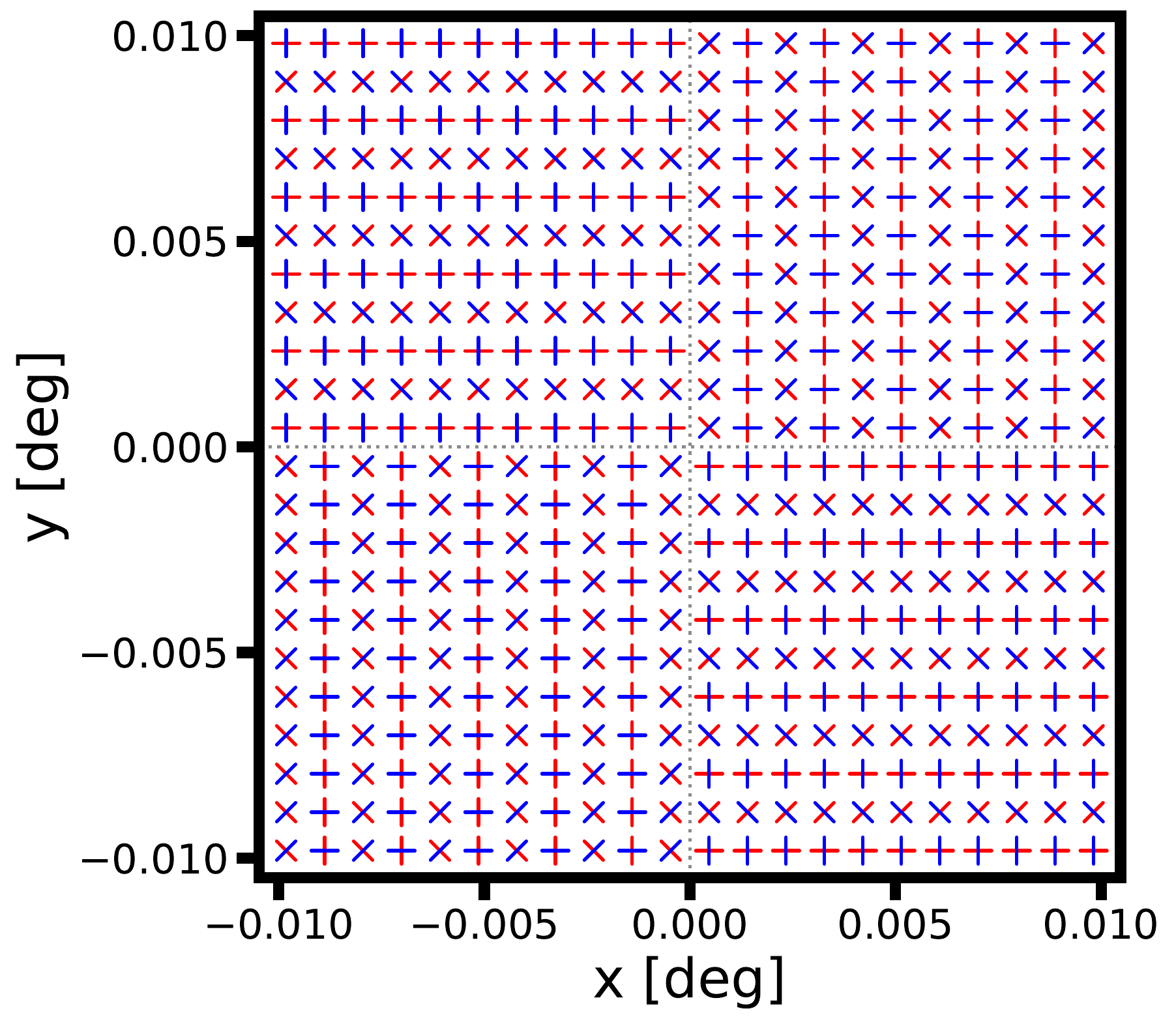}
\caption{The center of the focal plane model used in the simulations, projected on the sky. The focal plane is composed of 4 wafers (illustrated by the dividing dashed gray lines) in which detectors within a pair (top and bottom, illustrated by the blue and red bars respectively) are on top of each other. The different marker-pair angles correspond to pairs belonging to the two polarization modes $Q$ and $U$ (forming ``$+$'' and ``$\times$''-like shapes in the figure respectively). The full focal plane extends to 60 cm on the side.}
\label{fig:focal_plane}
\end{figure}

For the instrument specifications, instead of simulating a full-sized SO-like experiment, which could be a very numerically expensive task, we consider an instrument with 6,272 bolometers (3,136 detector pairs) distributed over 4 different detector wafers. The way in which detectors are wired in the focal plane, and the specific readout technology used in experiments, affect the electronic crosstalk systematic. We consider two hardware configurations based on $\fmux$ and $\mmux$ technologies, which we describe in more detail in Subsec.~\ref{sec:crosstalk}. The square focal plane we consider is 60 cm on the side and has a field of view (FOV) on the sky of $3\degree$\footnote{This makes up a subset of the full SO focal plane, which has a field of view of $\sim 5\degree$~\cite{Gudmundsson:2020wja}.}. The central region of the focal plane is shown in Fig.~\ref{fig:focal_plane}. Although the total number of detectors and the FOV are reduced compared to the current SO design, the configuration is the ballpark expected for CMB-focused frequency channels of large-aperture telescopes targeting CMB lensing surveys in the upcoming years. In the absence of specific pointing or polarization angle systematics, which we describe further below, each bolometer pair (top and bottom detectors in the following) in the focal plane has a specific coordinate such that two detectors within a pair are on top of each other and have a 90-degree difference in their polarization angle orientation. The focal plane is cut into four quadrants which represent a wafer. Within a quadrant, pixels form rows or columns which correspond to either $Q$ or $U$ modes in detector coordinates (with a fixed exact 45-degree difference between them, in absence of polarization angle systematics), depending on their polarization angle (indicated by the angle of the markers in the figure). Each quadrant is rotated by $90\degree$ with respect to the next quadrant. This layout is commonly adopted in the design of bolometric focal planes to allow an efficient averaging over orientation of angles during the scans.

All other instrument properties we consider are based on the SO `baseline' large aperture telescope (LAT) specifications at 145 GHz, described in Refs.~\cite{Hill:2018rva,Ade:2018sbj}. We use a baseline circularly-symmetric (CS) Gaussian beam with a full width at half maximum (FWHM) of 1.4 arcminutes. We rescale the SO baseline noise level to mimic the observations of an SO-like instrument scanning 5\% of the sky with a 20\% observation efficiency for 2.5 years. The corresponding map-domain white noise is 5.4 (7.6) $\mu K$-arcminutes for temperature (polarization) after 12-days of simulated scanning. Since the noise in the map is inhomogeneous due to the nature of the scan, this white-noise level is estimated from the power spectrum of weighted temperature and polarization noise maps. It is between the homogeneous noise level expected for an SO-like experiment observing 5\% of the sky for 5 years (3.5 $\mu K$-arcminutes), and the $10 \mu K$-arcminutes on 40\% of the sky expected for the baseline SO survey\footnote{These estimated values were obtained using the SO noise calculator, available at \href{https://github.com/simonsobs/so\_noise\_models}{https://github.com/simonsobs/so\_noise\_models}~\cite{Ade:2018sbj}, and assume a 20\% efficiency in the observing time.}. While using CMB maps with relatively low scaled noise may reduce noise-coupled systematic biases, the relative noise-related errors will also be smaller in our analysis, so we should still be sensitive to important effects.

Consistent with the design of SO LAT, our instrument model does not simulate the effects of a half-wave plate (HWP). Although a HWP could help to mitigate instrumental systematics for polarization~\cite{2010SPIE.7741E..2BB,Kusaka:2013pla}, especially if operated at cryogenic temperatures, the large-aperture telescopes typically used for lensing surveys do not normally use one as it is challenging to produce the large-sized plates required, and the HWP could also produce large unwanted systematics of its own~\cite{DAlessandro:2019snm}.

The configurations described above are used for all of our simulations. For each ``data'' set (apart from the systematics-free ones) we also include one systematic effect. Below, we describe how each injected systematic effect is modeled and how it affects the TOD. The impact of these systematics on the corresponding CMB power spectra and lensing reconstruction are discussed in Sec.~\ref{sec:systematics_biases}.

\subsection{Beam ellipticity}
\label{sec:beam_ellipticity}

An ideal bolometer observes a patch of the sky with a known shape (or beam), usually taken to be a circularly-symmetric (CS) Gaussian. Realistically, however, a detector's beam has some deviation from this symmetric shape. A realistic beam instead has an approximately elliptical shape with unequal minor and major axes which have some tilt angle with respect to the predefined focal plane axes. This means that realistic detectors do not observe the same sky area that detectors with circular beams would.
When each beam in an array of detectors has some different deviation from a CS shape, this can cause ``smearing'' effects in the resulting sky map. This could look similar to shearing or varying magnification expected from CMB lensing, and hence cause a lensing bias. When producing polarization maps using pair differencing, if the two beams of a detector-pair have a different shape there could be a substantial leakage between temperature and polarization measurements. This would induce biases both in the maps' power spectra and in the reconstructed lensing potential.

The effects of beam ellipticity on lensing reconstruction have been previously explored analytically~\cite{Miller:2008zi}, and several methods have been developed for mitigating beam asymmetry effects in CMB maps~\cite{Wallis:2014sha,Hivon:2016qyw}. Precise simulations of beam asymmetry in TOD simulations can be a very numerically-expensive task. The map of each simulated observation sample would need to be convolved with a specific pointing-dependent beam over a $4\pi$ solid angle~\cite{Mitra:2010rt}. Since it is too expensive to perform such convolution on a large number of samples (in our case, this would be performing a convolution over $4\times 10^{10}$ times for each simulation), we consider an approximate treatment for simulating the beam-shape systematic effects as TOD leakage terms following Ref.~\cite{Ade:2015fpw}. This approach does not account for the effect of far sidelobes, but these are expected to be more important for large-scale CMB modes that only have a minor impact on lensing reconstruction.

Given a temperature sample $T$ in a specific (time-dependent) position on the sky $\vecx$ and its corresponding beam $b(\vecx)$, which is not necessarily circular, the observed signal of this sample is
\begin{equation}
\begin{split}
\label{eq:T_obs}
T_{\obs}(\vecx) \equiv b(\vecx)\circledast T(\vecx) = \int b(\vecx-\vecy)T(\vecy)d\vecy . \\
\end{split}
\end{equation}
We can approximate the true beam $b(\vecx)$ as a perturbed CS Gaussian beam $b_{\CS}(\vecx-\vecy)$ with width $\sigma_{{\scriptstyle \FWHM}}$,
\begin{equation}
b(\vecx) \approx \alphaco{0}{}{}b_{\CS}(\vecx) + \alphaco{1,i}{}{}\frac{\partial b_{\CS}(\vecx)}{\partial x^i}+ \alphaco{2,ij}{}{}\frac{\partial^2b_{\CS}(\vecx)}{\partial x^i \partial x^j},
\label{eq:beam_expansion}
\end{equation}
where ${\alphaco{i}{}{}}$ are sets of expansion coefficients
for the $0^{\rm th}$-, $1^{\rm st}$- and $2^{\rm nd}$-order derivatives of $b_{\CS}(\vecx)$. Eq.~\eqref{eq:T_obs} is then approximated as
\begin{equation}
\label{eq:beam_leakages}
\begin{split}
&\begin{array}{rlll}
T_{\obs}(\vecx) & \approx \displaystyle\int \bigg[ &\alphaco{0}{}{}b_{\CS}(\vecx-\vecy) + {\alphaco{1,i}{}{}}\dfrac{\partial b_{\CS}(\vecx-\vecy)}{\partial x^i} \\
&& + {\alphaco{2,ij}{}{}}\dfrac{\partial^2 b_{\CS}(\vecx-\vecy)}{\partial
x^i \partial x^j} \bigg] T(\vecy) d\vecy
\end{array}\\
&\begin{array}{rl}
{\textcolor{white}{T_{\obs}(\vecx)}}&= \alphaco{0}{}{}T_{\beamed}(\vecx)+ {\alphaco{1,i}{}{}}\dfrac{\partial T_{\beamed}(\vecx)}{\partial x^i} + {\alphaco{2,ij}{}{}}\dfrac{\partial^2T_{\beamed}(\vecx)}{\partial x^i \partial x^j}
\end{array},
\end{split}
\end{equation}
where
\begin{equation}
\label{eq:beamed_map}
T_{\beamed}(\vecx) \equiv \int b_{\CS}(\vecx-\vecy)T(\vecy) d\vecy
\end{equation}
is the temperature signal convolved with the CS beam. We can therefore approximate the observed samples as a map convolved with a CS beam, $T_{\beamed}$, plus leakage terms as shown in Eq.~\eqref{eq:beam_leakages}. The leakage terms depend on the derivatives of $T_{\beamed}$ and on the coefficients $\alphaco{i}{}{}$ which are derived from expanding the perturbed beam $b(\vecx)$ around the CS beam $b_{\CS}$. Instead of repeating this convolution process for each sample, we can obtain $T_{\beamed}$ from convolving our input sky map with the CS beam and use this map and its derivatives to get the leakage terms for each observed sample.

Using this treatment, we can analyze how these leakage terms affect the TOD and the resulting temperature and polarization signals. The time streams of top and bottom detectors within a pair (two orthogonal detectors which are in this case aimed towards the same sky area) can be written as
\begin{equation}
\begin{split}
d_{\t} &= b_{\t} \circledast \left[T + Q\cos{\left(2\psi\right)} + U\sin{\left(2\psi\right)}\right], \\
d_{\b} &= b_{\b} \circledast \left[T - Q\cos{\left(2\psi\right)} - U\sin{\left(2\psi\right)}\right],
\end{split}
\end{equation}
where $b_{\t}$ and $b_{\b}$ are the top and bottom bolometers' beams, respectively, and $\psi$ is the polarization angle. The temperature and polarization time streams are then given by the sum and the difference of the pair's time streams:
\begin{equation}
\begin{split}
d_{+} &= b_{+} \circledast T + b_{-} \circledast \left[Q\cos{\left(2\psi\right)} + U\sin{\left(2\psi\right)}\right], \\
d_{-} &= b_{-} \circledast T + b_{+} \circledast \left[Q\cos{\left(2\psi\right)} + U\sin{\left(2\psi\right)}\right],
\label{eq:d_minus_plus}
\end{split}
\end{equation}
where
\begin{equation}
b_{\pm} \equiv \dfrac{b_{\t} \pm b_{\b}}{2}.
\end{equation}
Repeating the beam approximation above for the convolution terms in Eq.~\eqref{eq:d_minus_plus}, we get
\begin{equation}
\begin{split}
d_{+} &= \alphaco{0(+)}{}{}T_{\beamed} + {\alphaco{1,i(+)}{}{}}\dfrac{\partial T_{\beamed}(\vecx)}{\partial x^i} + {\alphaco{2,ij(+)}{}{}}\dfrac{\partial^2T_{\beamed}(\vecx)}{\partial x^i \partial x^j} \\
&+ \alphaco{0(-)}{}{}P_{\beamed}+{\alphaco{1,i(-)}{}{}}\dfrac{\partial P_{\beamed}(\vecx)}{\partial x^i} + {\alphaco{2,ij(-)}{}{}}\dfrac{\partial^2P_{\beamed}(\vecx)}{\partial x^i \partial x^j}, \\
d_{-} &= \alphaco{0(-)}{}{}T_{\beamed} + {\alphaco{1,i(-)}{}{}}\dfrac{\partial T_{\beamed}(\vecx)}{\partial x^i} + {\alphaco{2,ij(-)}{}{}}\dfrac{\partial^2 T_{\beamed}(\vecx)}{\partial x^i \partial x^j} \\
&+ \alphaco{0(+){}{}}P_{\beamed}+{\alphaco{1,i(+)}{}{}}\dfrac{\partial P_{\beamed}(\vecx)}{\partial x^i} + {\alphaco{2,ij(+)}{}{}}\dfrac{\partial^2P_{\beamed}(\vecx)}{\partial x^i \partial x^j},
\label{eq:d_minus_plus_explicit}
\end{split}
\end{equation}
where the coefficients ${\alphaco{i(\pm)}{}{}}$ correspond to $b_{\pm}$, and we define $P_{\beamed} \equiv Q_{\beamed}\cos{\left(2\psi\right)} + U_{\beamed}\sin{\left(2\psi\right)}$, the polarization field convolved with the CS beam in analogy with Eq.~\eqref{eq:beamed_map}, for convenience. In practice, the ${\alphaco{i(\pm)}{}{}}$ coefficients are time-dependent. The time dependency is due to the different orientation of the expansion basis used for computing their values and the sky coordinate system at a given observation time. The difference in orientation can be easily accounted for by rotating the coefficients computed in Eq.~\eqref{eq:beam_expansion} by a suitable angle.

\begin{figure}[!h]
\centering
\hspace{-0.03\textwidth}
\subfloat[CS beams.]{\includegraphics[width=0.34\columnwidth]{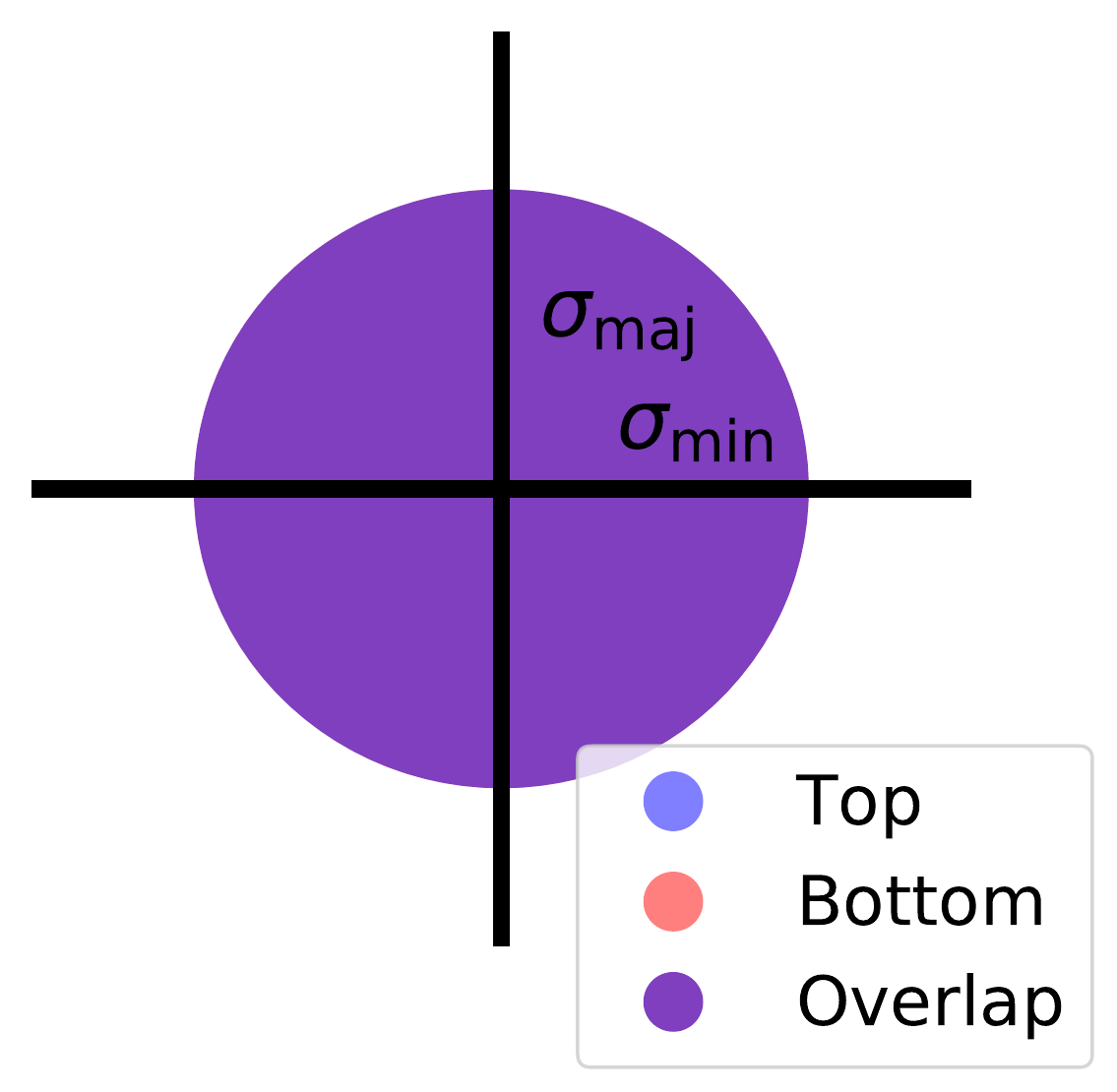}\label{fig:CS_beams}}
\subfloat[Differential beam ellipticities.]{\includegraphics[width=0.34\columnwidth]{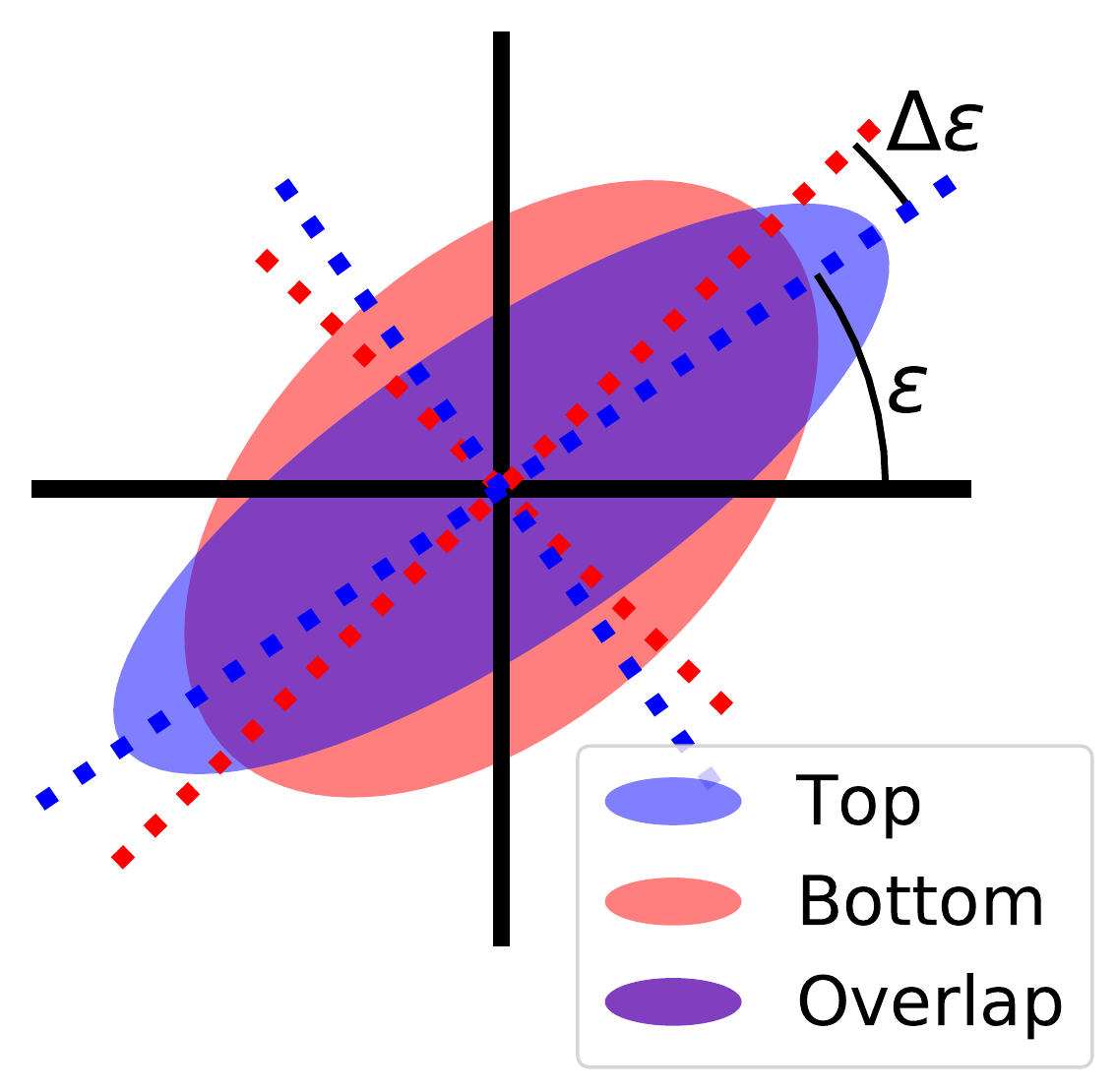}\label{fig:differential_beam_ellipticities}}
\subfloat[Differential pointing.]{\includegraphics[width=0.34\columnwidth]{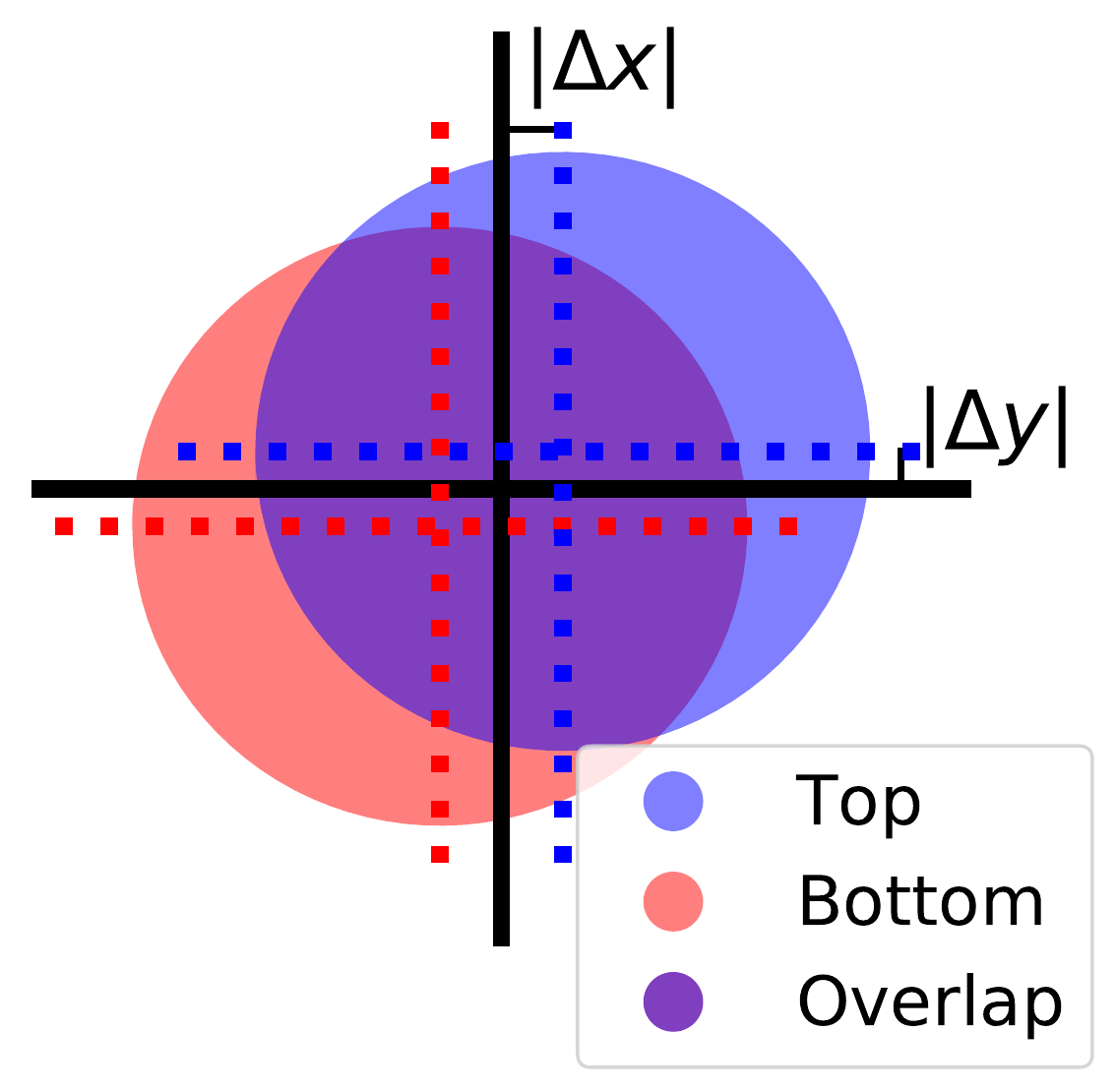}\label{fig:differential_pointing}}
\caption{Illustration of the beam models and parameters in Eq.~\eqref{eq:beams}. Panel (a) shows the two overlapping circularly-symmetric (CS) Gaussian beams ($\sigma_{\maj}=\sigma_{\min}=\sigma_{{\scriptstyle \CS}}$) of a detector pair. This represents the beam shape in all simulations apart from the beam ellipticity and differential pointing ``data'' simulation sets. Panel (b) shows the beam ellipticity model we use. In this case, the two beams of a detector pair have a different ellipticity: $\sigma_{\maj}\neq\sigma_{\min}$ for each beam, different axes lengths for each detector, and some angle difference $\Delta\varepsilon$ also exists between their major axes (on top of the $90\degree$ orthogonality of the two detectors). Panel (c) shows the same unperturbed beam shapes as in (a), however in this case each beam's center is shifted. This is our differential pointing model, in which the beam centers of a detector pair are shifted according to Eq.~\eqref{eq:diffp}.}
\label{fig:beam_ellipticity}
\end{figure}

Leakage which results from $\alphaco{0(+)}{}{} \neq 1$, and thus from a loss of optical power, is usually mitigated during gain calibration or polarization efficiency estimation. We therefore set $\alphaco{0(+)}{}{}$ to $1$ for all detectors to focus on the less trivial leakage terms, and perform a separate analysis of gain systematics in a later subsection. When a pair's beams have the same shape, even if elliptical, $b_{-}=0$ and $b_{+}=b_{\t}=b_{\b}$. In this case, $\alphacovec{1(-)}{}{}=\alphacovec{2(-)}{}{}=0$, and all $T \rightarrow P$ and $P \rightarrow T$ leakage terms vanish. Any biases in this scenario are attributed only to $T \rightarrow T$ and $P \rightarrow P$ leakages which depend on the deviation of the elliptical beam $b_{+}$ from being CS. When a pair's beams do not have the same shape, the $T \rightarrow P$ and $P \rightarrow T$ leakage terms do not vanish. This can induce a significant bias in the polarization maps due to the large temperature signal amplitude, which can then affect the lensing reconstruction. We simulate the most general case of beam asymmetry systematics described above, in which all leakage terms (apart from the gain-related ones) are injected into the TOD. To calculate the fitting coefficients ${\alphaco{i(\pm)}{}{}}$, we define the CS and the parameterized elliptical beams as
\begin{eqnarray}
b_{\CS}(\vecx) &\equiv& \frac{1}{2\pi\sigma_{{\scriptstyle \CS}}^2} e^{-\frac{\vecx^2}{2\sigma_{{\scriptstyle \CS}}^2}} , \nonumber\\[2pt]
b(\vecx) &\equiv& \frac{1}{2\pi\sigma_{\min}\sigma_{\maj}} e^{-\frac{1}{2}\left[\sigmavec^{-1}\cdot\Rvec(\varepsilon)\cdot\vecx\right]^2},
\label{eq:beams}
\end{eqnarray}
where
\begin{eqnarray}
\sigma_{{\scriptstyle \CS}} &\equiv& \frac{\sigma_{{\rm \FWHM}}}{\sqrt{8\ln{2}}}, \nonumber\\
\sigmavec &\equiv&
\left(
\begin{array}{cc}
\sigma_{\maj}	& 0					\vspace{0.05in}\\
0					& \sigma_{\min}
\end{array}
\right) , \nonumber\\
\Rvec(\varepsilon) &\equiv&
\left(
\begin{array}{rr}
\cos(2\varepsilon) & -\sin(2\varepsilon) \vspace{0.05in}\\
\sin(2\varepsilon) & \cos(2\varepsilon)
\end{array}
\right),
\end{eqnarray}
$\sigma_{\maj}$ ($\sigma_{\min}$) is the size of the semi-major (minor) axis, and $\Rvec$ is a matrix responsible for rotating the ellipse by some angle $\varepsilon$ between the major axis of beam ellipse and the focal plane's $x$ axis. The beam parameters are illustrated in Fig.~\ref{fig:beam_ellipticity}.
The minor and major ellipse axes of each beam deviate symmetrically from $\sigma_{{\scriptstyle \CS}}$,
\begin{equation}
\sigma_{\maj \atop \min} = \sigma_{\scriptstyle \CS} \pm \frac{\Delta\sigma}{2},
\label{eq:perturbed_beam}
\end{equation}
where $\Delta\sigma$ is determined using the ellipticity $e_{\beam}$ definition,
\begin{equation}
e_{\beam} \equiv \frac{\sigma_{\maj}^2 - \sigma_{\min}^2}{\sigma_{\maj}^2 + \sigma_{\min}^2}.
\end{equation}

\begin{figure}[!h]
\centering
\includegraphics[width=\columnwidth]{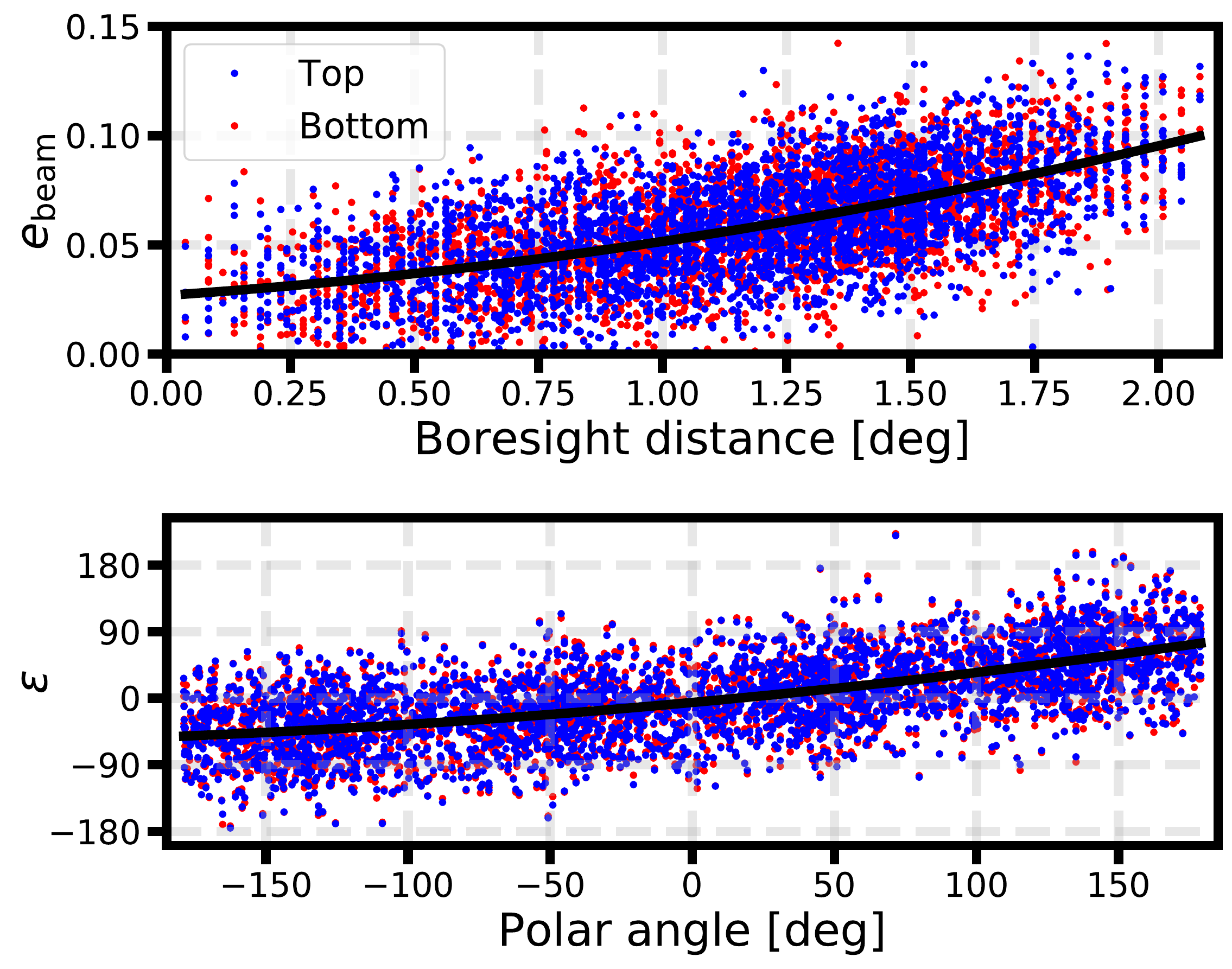}
\caption{Beam ellipticities and angles of all simulated top (blue markers) and bottom (red markers) detectors. Each plot shows the function used as the normal distribution mean for generating the beam random variables (black lines).}
\label{fig:correlated_beam_ellipticities_and_angles}
\end{figure}

Each detector beam is assigned with a random ellipticity $e_{\beam}$ and a random ellipticity angle $\varepsilon$. To add a level of realism to the ellipticity models, we correlate each ellipticity and angle to the detector's distance from the boresight coordinates and polar angle, respectively, by assuming a $2^{\rm nd}$ degree polynomial that mimics the fact that detectors observing regions close to the edge of the FOV are subject to more optical distortions\footnote{A good demonstration of these correlations is shown in Ref.~\cite{Gudmundsson:2020wja}, where a more comprehensive review of the SO optics can also be found.}. The polynomial functions, along with the beam ellipticities and angles, are shown in Fig.~\ref{fig:correlated_beam_ellipticities_and_angles}. The ellipticities and angles are drawn from a normal distribution with mean centered on the respective polynomial function with 2\% and $45\degree$ standard deviations, respectively. These dispersion values are consistent with e.g. POLARBEAR~\cite{Ade:2017uvt} and BICEP2~\cite{2019ApJ...884..114B} beam measurements. On top of the ellipticity angles, which are the same for two detectors in a pair, a random differential angle $\Delta\varepsilon$ is also used to perturb the beams of all bottom detectors. These angles are drawn from a normal distribution with a zero mean and a $5\degree$ width. All beam parameters are drawn once per simulation and therefore remain constant in time throughout the simulated observation period. The relevant derivatives of the temperature and polarization maps, which are used in the leakage terms, are obtained using the \texttt{synfast} routine of the HEALPix~\cite{Gorski:2004by} package.

In our simulations, we do not model the cross-polar beam response. This response is expected to be subdominant for an SO-like instrument based on modern optical coupling technologies for bolometric detectors and cross-Dragone telescopes~\cite{Crowley:2018eib,Gallardo:2018rix}. We also assume that all baseline beams have a perfect circular shape with a Gaussian radial profile. In practice, this is an approximation, as diffraction effects in the optics will cause the beam to decay asymptotically as $\sim 1/\theta^3$, where $\theta$ is the angle from the beam peak~\cite{Hasselfield:2013zza,Ade:2017uvt}. Any characterization of the beam properties in the field through dedicated calibration observations will naturally include these effects in the main beam model, and hence include it in the transfer function used for subsequent steps of the data analysis. While we do not include diffraction effects in our baseline beams, we found that when perturbing our elliptical beams around a beam which includes diffraction tails, the resulting leakage coefficients ${\alphaco{i(\pm)}{}{}}$ are similar to those obtained using the fully Gaussian beam. As such, although the diffraction tails affect the beam beyond its FWHM scale, we do not expect these corrections to significantly change our results for beam-related or other systematics considered in this work. In general, diffraction tails could be important, as they allow the telescope to pick up spurious emissions coming from the ground or other astronomical sources, and would have to be included in a dedicated analysis of the telescope sidelobes, which we did not consider in this work.

The $b_{\pm}$ maps are shown in Fig.~\ref{fig:beam_maps} for a detector pair from our simulations. The beam difference $b_{-}$ map for beam ellipticity has quadrupole-like symmetry, corresponding to the main leakage terms for this systematic coming from the $2^{\rm nd}$ derivative terms. The map-level effects of beam ellipticity for temperature and polarization are shown in Fig.~\ref{fig:syst_maps}. Compared to the simulation with CS beams, the residuals seem negligible relative to the signal amplitudes. Expectedly, the polarization residuals are larger than those of the temperature due to the $T \rightarrow P$ leakage terms.

\subsection{Differential pointing}
\label{sec:differential_pointing}

Another beam-related systematic that we model results from detectors in a focal plane pixel not being centered on the same sky coordinates. In other words, the beams of two detectors in a pair are not aligned in the focal plane reference frame. When this occurs, the temperature and polarization maps, which are produced from the sum and difference of the pair data streams, will be distorted. Typically, this is mitigated during map-making by considering the mid-point beam centers as the true center of each beam in a pair. The residual of this effect can, however, produce smearing features and $T\rightarrow P$ leakage in the maps, which could potentially propagate to the lensing reconstruction. Analytic approaches for characterizing the differential pointing effects on the lensing potential were previously explored in Refs.~\cite{Miller:2008zi,Su:2009tp}.

In our simulations, this systematic effect is modeled by introducing an offset to the beam-center coordinates of two detectors within a pair. A different offset is drawn for each detector pair. For a given pair, the unperturbed pointing coordinates $(x_0,y_0)$ in the focal plane reference frame is shifted by
\begin{equation}
(\Delta x,\Delta y)_{{\scriptstyle \t \atop \scriptstyle \b}} = \pm \frac{\rho}{2}\left(\cos{\theta}, \sin{\theta}\right),
\label{eq:diffp}
\end{equation}
where $\rho \in \N(15'',1.5'')$ is the offset magnitude and $\theta \in \U(0,2\pi)$ is the offset direction angle with respect to the horizontal focal plane axis. The magnitude of the $\rho$ that we use is conservative, as current-generation experiments with on-chip detectors achieved differential pointing well below the mean value assumed here~\cite{pb2014}. The differential pointing offset is illustrated in Fig.~\ref{fig:differential_pointing}. Following the previous sections, the perturbed beams are then used to compute the coefficients ${\alphaco{i}{}{}}$ for the leakage terms which are injected into the temperature and polarization time streams as in Eq.~\eqref{eq:d_minus_plus_explicit}.

\begin{figure}[!ht]
\centering
\includegraphics[width=\columnwidth]{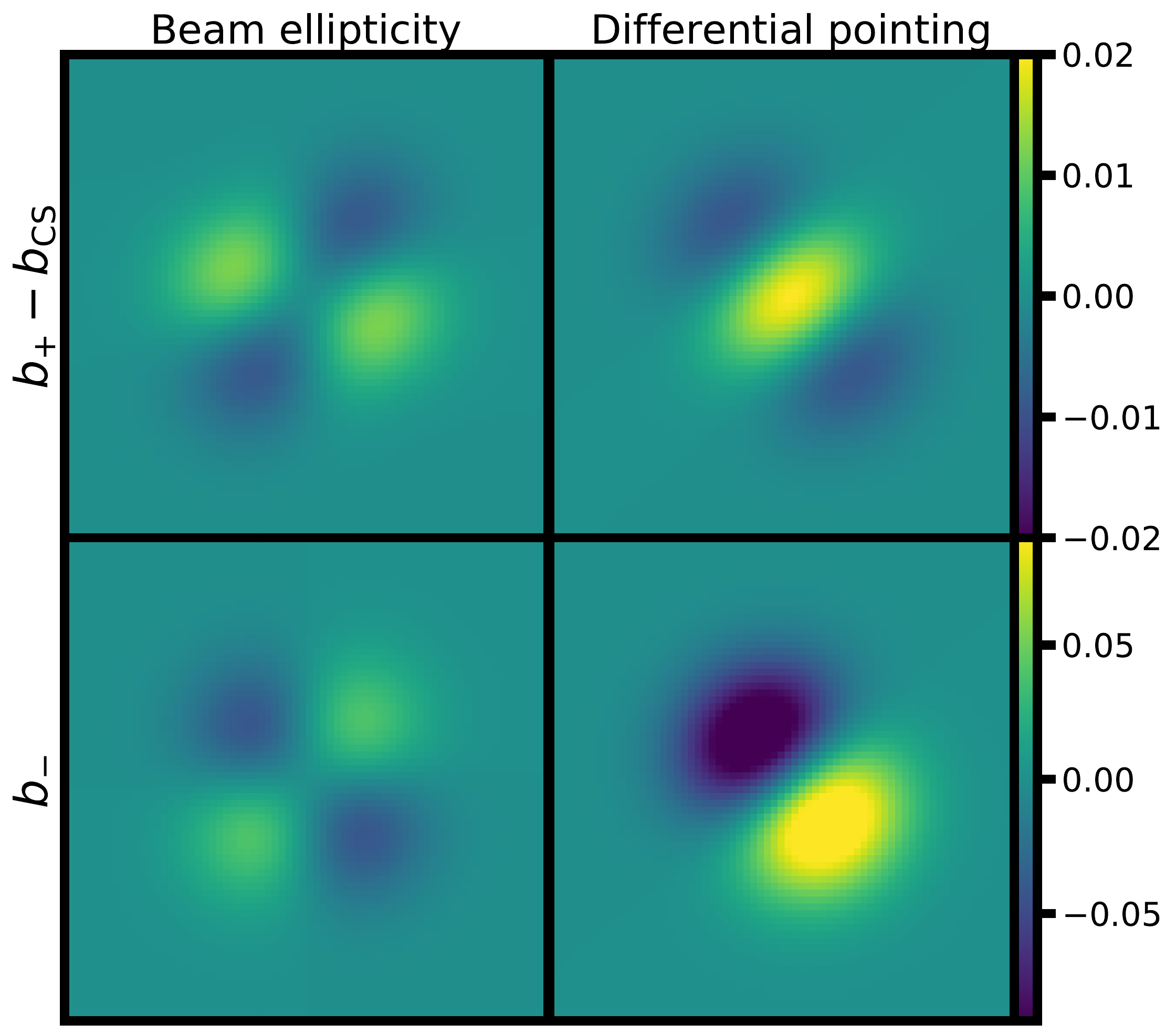}
\caption{
\textit{Row 1}: Difference between $b_{+}$ maps of the perturbed and CS beams for the beam ellipticity (left panel) and differential pointing (right panel) systematics for a detector pair in our simulations. The deviation from CS is larger for the differential pointing systematic.
\textit{Row 2}: The $b_{-}$ maps for the beam ellipticity (left panel) and differential pointing (right panel) systematics of a detector pair from our simulations. As with $b_{+}$, the difference between the two beams within a pair is larger for the differential pointing systematic. The shape of $b_{-}$ for beam ellipticity (differential pointing) has quadrupole- (dipole-)like features.}
\label{fig:beam_maps}
\end{figure}

The $b_{\pm}$ maps for the differential pointing systematic are shown in Fig.~\ref{fig:beam_maps} for a detector pair from our simulations. The beam difference $b_{-}$ map has dipole-like features, which suggests that the main leakage terms for this systematic would stem from the $1^{\rm st}$ derivative leakage terms.

\begin{figure}[!h]
\includegraphics[width=0.975\columnwidth]{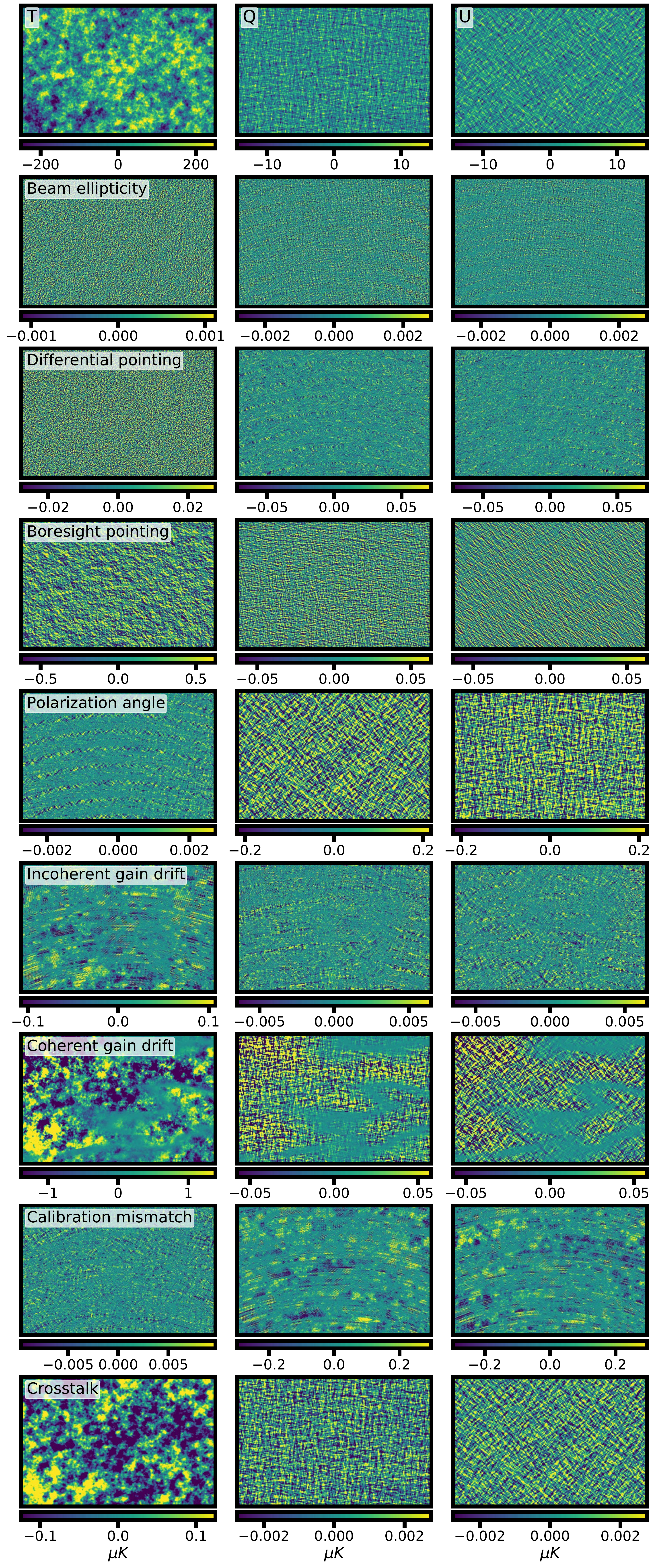}
\caption{\textit{Row 1}: White-noise-free temperature (left), $Q$ (middle) and $U$ (right) maps of a systematics-free ``data'' simulation.
\textit{Rows 2-8}: Difference maps of the same realization with and without systematics for temperature (left column), $Q$ (middle column) and $U$ (right column).
All panels show the middle area of the full simulated box with dimensions $17\times25.5 \deg^2$.}
\label{fig:syst_maps}
\end{figure}

The map-level effects of differential pointing for temperature and polarization are shown in Fig.~\ref{fig:syst_maps}. The differential pointing residuals for both temperature and polarization maps are larger compared to the beam ellipticity residuals. For both of these beam-related systematic effects, the temperature residuals appear to be nearly spatially uncorrelated. The polarization residuals in the differential pointing case do not have $Q$- and $U$-like features as with the beam ellipticity residuals.

\subsection{Boresight pointing}
\label{sec:boresight_pointing}

The systematics discussed in the previous subsections involved perturbed models of detector beams. In this subsection, we describe a systematic effect that is produced from inaccuracies in the pointing coordinates of the entire focal plane, or \textit{boresight pointing}, during scans. The exact pointing of the telescope needs to be reconstructed from the position of known sources. The direction in which the focal plane of a telescope is pointing during a scan might differ slightly from the pointing direction recorded by the telescope position encoders. These errors can originate from wind gusts, temperature changes, temperature gradients across the focal plane due to heating of the telescope structure, vibrations due to the motion of the telescope, deformation of the telescope's mirror due to its own weight, and more. The errors due to deformation can mostly be corrected by estimating the variations of the pointing correction (that relates the recorded telescope position to the position of known sources) as a function of time, while other effects can be assumed as random. We therefore simulate the pointing errors by perturbing the boresight's azimuth and elevation for each sampling, while using their original unperturbed values in the pointing matrix in the map-making stage. The azimuth and elevation offsets are drawn from a normal distribution with a 3 arcseconds mean (a typical precision of a telescope position encoder) and a variance such that the total pointing uncertainty is 13 arcseconds and error in azimuth and elevation are uncorrelated. This pointing error is about $\sim 10\%$ of the CS beam FWHM we considered in this work and is consistent with typical results of state-of-the-art experiments~\cite{pb2014}.

The map-level effects of the boresight pointing systematic for temperature and polarization are shown in Fig.~\ref{fig:syst_maps}. Unlike the previously mentioned systematics, the residuals from perturbing the boresight coordinates are as important for temperature as for polarization. The map-level residuals are small-scale changes arising from a small additional smoothing-like effect on the maps due to the randomized pointing. This systematic does not produce any $T \leftrightarrow P$ mixing.

\subsection{Polarization angle}
\label{sec:polarization_angle}

The accuracy of polarization angle measurements is important to correctly characterize the $E$ and $B$ modes of the CMB~\cite{2013ApJ...762L..23K,Abitbol:2020fvn}. If the true polarization angles of each detector deviate from their estimated values, which are used to make the $Q$ and $U$ maps, $E$/$B$ mixing is introduced. This not only contaminates the resulting $E$ and $B$ modes, but also produces non-zero $EB$ and $TB$ correlations. These correlations are expected to vanish in cosmological models where parity is preserved~\cite{PhysRevLett.78.2058,PhysRevD.55.7368,Zaldarriaga:1996xe,Lepora:1998ix,Lue:1998mq,Ferte:2014gja}. Models that include non-standard physical mechanisms which manifest on cosmological scales (such as cosmic birefringence, parity violation) predict the existence of intrinsic $EB$ or $TB$ correlations that can also get contaminated by a polarization angle miscalibration~\cite{Pagano:2009kj}.

In general, the polarization angles of the top and bottom detectors can each be different from the expected angle $\psi$ by a different $\Delta\psi$, such that the time streams of a detector pair are
\begin{eqnarray}
d_{\t} &= T &+ Q\cos{\left[2\left(\psi + \Delta\psi_{\t}\right)\right]} \nonumber\\
&&+ U\sin{\left[2\left(\psi + \Delta\psi_{\t}\right)\right]}, \nonumber\\
d_{\b} &= T &- Q\cos{\left[2\left(\psi + \Delta\psi_{\b}\right)\right]} \nonumber\\
&& - U\sin{\left[2\left(\psi + \Delta\psi_{\b}\right)\right]}.
\end{eqnarray}
The temperature and polarization time streams are then
\begin{eqnarray}
d_{+} = T &+ &Q\frac{\cos{\left[2\left(\psi + \Delta\psi_{\t}\right)\right]} - \cos{\left[2\left(\psi + \Delta\psi_{\b}\right)\right]}}{2} \nonumber\\
&+ &U\frac{\sin{\left[2\left(\psi + \Delta\psi_{\t}\right)\right]} - \sin{\left[2\left(\psi + \Delta\psi_{\b}\right)\right]}}{2}, \nonumber\\
d_{-} = &&Q\frac{\cos{\left[2\left(\psi + \Delta\psi_{\t}\right)\right]} + \cos{\left[2\left(\psi + \Delta\psi_{\b}\right)\right]}}{2} \nonumber\\
&+ &U\frac{\sin{\left[2\left(\psi + \Delta\psi_{\t}\right)\right]} + \sin{\left[2\left(\psi + \Delta\psi_{\b}\right)\right]}}{2}.\nonumber\\
\label{eq:d_minus_plus_polang}
\end{eqnarray}
When $\Delta\psi_{\t}=\Delta\psi_{\b} \equiv \Delta\psi$, there is no $P \rightarrow T$ leakage. When all detector pairs are perturbed with the same $\Delta\psi$ value, the recovered polarization maps are effectively equivalent to the true polarization sky signals rotated by a constant angle $\Delta\psi$. Under this assumption, it is straight-forward to propagate this systematic effect all the way to the $E$ and $B$ modes and their power spectra~\cite{2013ApJ...762L..23K}. When the polarization angle perturbations in the top and bottom detectors are completely anti-correlated, i.e. $\Delta\psi_{\t}=-\Delta\psi_{\b}$, the polarization bias becomes an effective gain error of the size $\cos(2\Delta\psi_{\t})$. In this work, we consider the most general case, in which both $\Delta\psi_{\t}$ and $\Delta\psi_{\b}$ are independently drawn for each detector pair, such that both the polarization and temperature signals are affected by the perturbed angles.

We model differential polarization angle perturbations by drawing a different value for $\Delta\psi$ from $\N(-1.1\degree,0.5\degree)$ for each detector in each detector pair. This perturbation level will be referred to as ``setup A'' throughout the paper. The perturbation values for this setup are consistent with the polarization angle errors measured by POLARBEAR and BICEP2 prior to applying a polarization angle self-calibration procedure (see~\cite{Ade:2014afa,bicep22014} and discussion around Eq.~\eqref{eq:pol_ang_powers} in Sec.~\ref{sec:mitigation_techniques}). Other experiments, such as ACTPol and SPTpol, have reported lower mean values for the polarization angle errors consistent with $\Delta\psi\sim 0.5 \degree$ \cite{Thornton:2016wjq,bianchini2020}. Although we do not consider this case in detail for estimating lensing biases, we give a comparison between setup A and an ACTPol-like setup with $\Delta\psi$ drawn from $\N(-0.5\degree,2.0\degree)$ (``setup B'') in Secs.~\ref{sec:lensing_power_biases} and~\ref{sec:mitigation_techniques}.

The map-level effects of the polarization angle systematic for temperature and polarization are shown in Fig.~\ref{fig:syst_maps}. The small temperature residuals are the $P \rightarrow T$ leakage induced by the differential polarization angles within a detector pair. These residuals also show the scanning strategy stripes due to its correlation with the polarization angles. Because this systematic effectively rotates the polarization maps, the $Q$ residual map mostly consist of $U$ features, and vice versa.

\subsection{Gain drifts}
\label{sec:gain_drifts}

In this and the next subsection we discuss detector-level systematics which relate to the TOD gains. During an observation run, various internal or external factors could change the measured bolometer gain that calibrates the raw data to physical units: local temperature gradients across the focal plane could induce gain drifts for each detector pair until a gain recalibration is performed; external heating of the entire focal plane or a coherent change of the detectors' optical loading could cause a coherent gain drift for all detectors. In our simulations, we model these gain drift effects as a function of time using a linear drift model,
\begin{equation}
g(t) = 1 + \Delta g \frac{\left( t \mod t_R \right)}{t_R},
\label{eq:gain_drift}
\end{equation}
such that after each time interval $t_R$ the gain is recalibrated back to unity. This assumes that the gain calibration procedure restores a perfect calibration relative to the input map. As such, we assume potential effects due to bandpasses can be characterized with a sufficient level of precision. In Subsec.~\ref{sec:calibration_mismatch} we also consider a related effect in which the recalibration produces some gain mismatch between the detectors in a pair, producing an inter-calibration problem. We consider a retuning interval of $\sim1.2$ hours, and draw the gain perturbation variable $\Delta g$ for each pair (and once for each drifting duration) from a normal distribution with a zero mean and a 0.05 width. Although the retuning interval can be optimized depending on the exact scanning strategy, we use a value similar to those employed for observations performed from the Atacama plateau with a similar scanning strategy~\cite{Adachi:2019mjv}.

\begin{figure}[!h]
\centering
\includegraphics[width=\columnwidth]{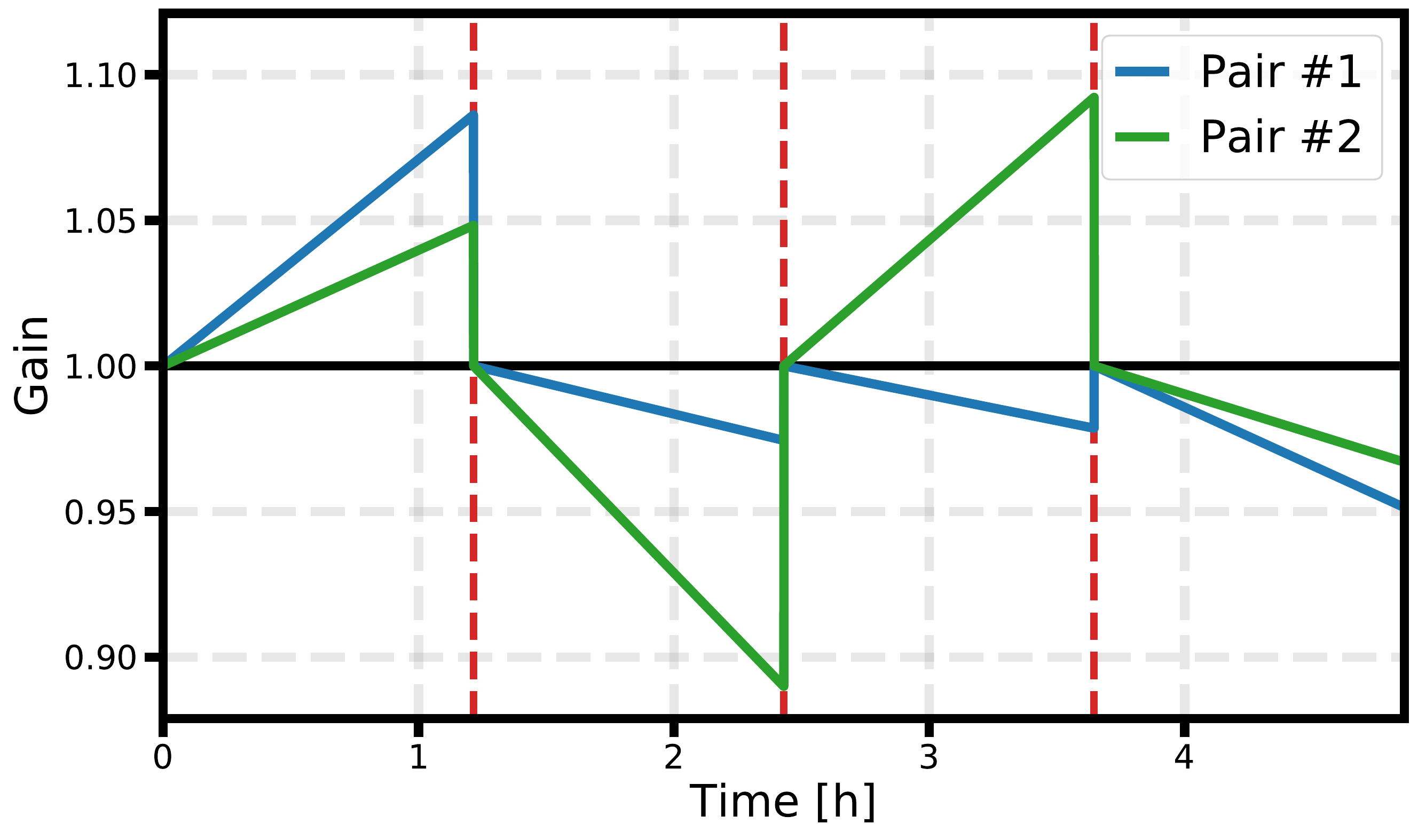}
\caption{Illustration of the linear gain drift model during the course of CES 1 from Fig.~\ref{fig:CES} for two detector pairs (green and blue lines). Recalibration occurs 4 times during each CES, shown by the dashed red lines. For this CES, which lasts $\sim5$ hours, recalibration occurs every $\sim1.25$ hours. For the shorter CESs, recalibration occurs roughly every hour. This illustration shows the incoherent gain drift of two detector pairs. In our coherent gain drift simulations, the green gains are identical to the blue gains, and represent the drift of all detector pairs.}
\label{fig:gain_drift}
\end{figure}

This gain drift model, which is illustrated in Fig.~\ref{fig:gain_drift}, simulates the effect of \textit{incoherent gain drifts} between detector pairs. As mentioned before, another possibility is a \textit{coherent gain drift} across the focal plane. We therefore produce an additional simulation set for this effect. We consider these two extreme gain drift scenarios instead of simulating a local (on the focal plane) drifting model to demonstrate how different drifting scenarios affect the lensing reconstruction. For coherent gain drift, the same random variable $\Delta g$ is used for all detector pairs for each drifting duration, but is different after each recalibration. This type of drift is similar to that illustrated in Fig.~\ref{fig:gain_drift}, but with the gains of both pairs (and all other pairs in the focal plane) being the same in each drifting period.

Since gain is a multiplicative parameter for the time-stream signal and noise,
\begin{equation}
\begin{split}
d_{\t} &= g\left[T + Q\cos{\left(2\psi\right)} + U\sin{\left(2\psi\right)}+n\right], \\
d_{\b} &= g\left[T - Q\cos{\left(2\psi\right)} - U\sin{\left(2\psi\right)}+n\right],
\label{eq:gain_drift_d}
\end{split}
\end{equation}
we can estimate its average effect on the 2-point and 4-point correlation functions, which we define later in Sec.~\ref{sec:lensing_analysis}, analytically.

The map-level effects of the incoherent and coherent gain drift systematics for temperature and polarization are shown in Fig.~\ref{fig:syst_maps}. Expectedly, the incoherent gain drift residuals are much smaller compared to the coherent drifts. This happens because when different pairs have a different drift, the systematic effect averages out quickly for a sky area which is observed by multiple pairs over time. Coherent drift biases take a longer time to average out. As such, their residuals leave large-scale areas which are affected by an incorrect calibration. The resulting patterns depend on the scanning strategy, sampling frequency, and gain recalibration frequency. Longer observations using the same basic scan, but a more frequent recalibration strategy, would change these patterns and reduce the residual amplitude.

\subsection{Calibration mismatch}
\label{sec:calibration_mismatch}

\begin{figure}[!h]
\centering
\includegraphics[width=\columnwidth]{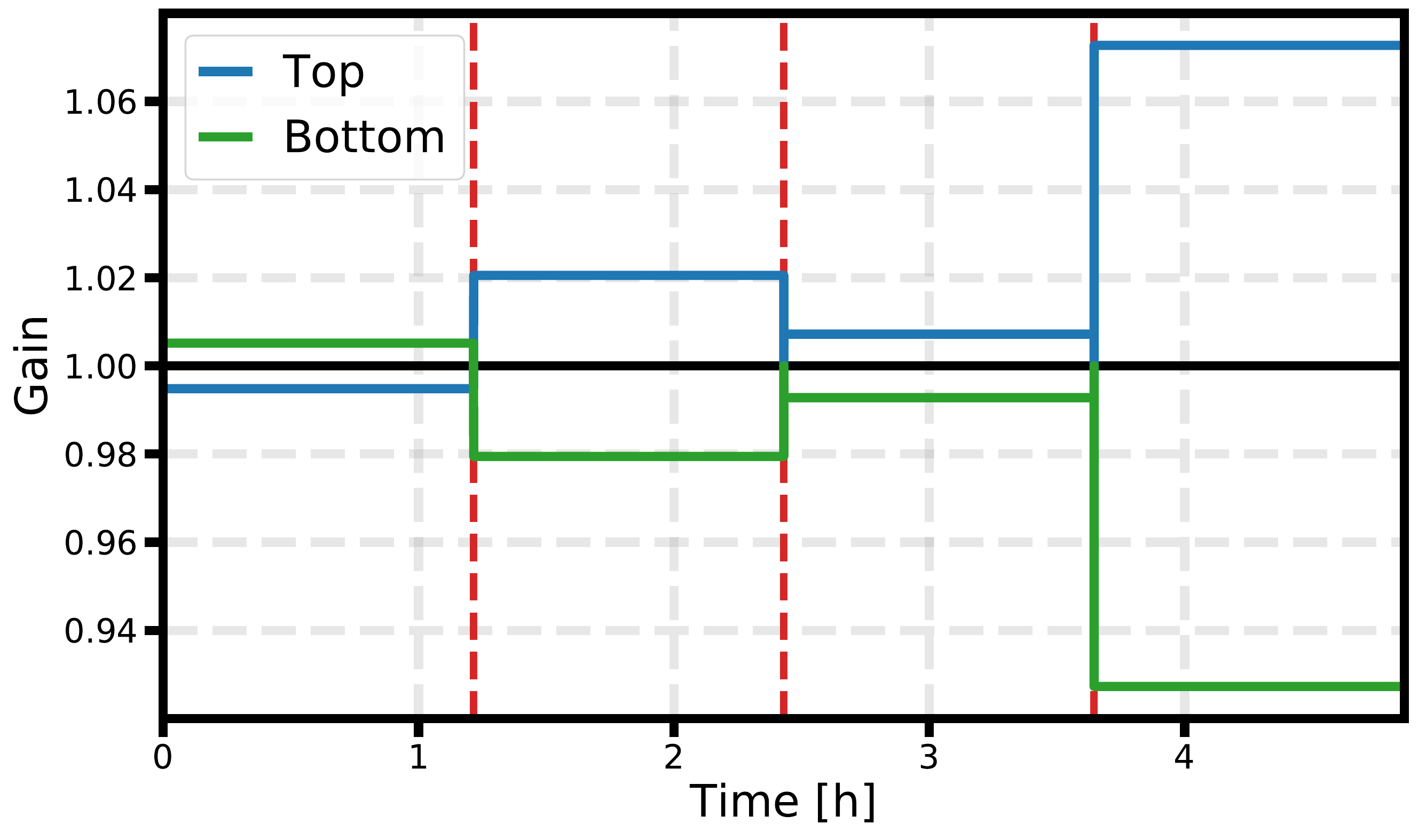}
\caption{Illustration of the calibration mismatch model during the course of CES 1 from Fig.~\ref{fig:CES} for top (blue) and bottom (green) detectors within a pair. As with the gain drift model which is illustrated in Fig.~\ref{fig:gain_drift}, recalibration occurs 4 times during each CES, shown by the dashed red lines. For this CES, which lasts $\sim5$ hours, recalibration occurs every $\sim1.25$ hours. For the shorter CESs, recalibration occurs roughly every hour. Unlike the gain drift model, here each calibration process adds a symmetrically gain distortion between the two detectors in a pair.}
\label{fig:calibration_mismatch}
\end{figure}

Another gain-related systematic effect results from an inaccurate gain inter-calibration process between two detectors in a given pair. During an observation run, gains are usually calibrated back to unity multiple times. This recalibration process could potentially produce some level of differential gain, or \textit{calibration mismatch}, if the new gains of a detector pair are not equal. In this case, each detector gain has some deviation from unity. This gain offset can be different after each gain recalibration. We simulate this effect by symmetrically offsetting the top and bottom gains $g$ of each detector pair such that
\begin{equation}
g_{\t}(t)-g_{\b}(t) = 2\epsilon_g(t).
\end{equation}
We model this systematic effect symmetrically so that only the leakage terms in the temperature and polarization time streams depend on the gain mismatch level, and the overall absolute calibration of the $T$, $Q$ and $U$ Stokes parameters is not affected. This is consistent with the choice made in the previous sections where we assumed absolute calibration and polarization efficiencies effects can be correctly measured or calibrated on other external data sets such as e.g. Planck. A different gain offset $\epsilon_g$ is applied after each calibration and for each detector pair. The modified gains of a detector over time are illustrated in Fig.~\ref{fig:calibration_mismatch}. We use the same probability distribution as before, $\N(0,0.05)$, to draw a different offset $\epsilon_g$ for each detector pair. This distribution is a conservative estimate of possible gain systematics as current generation of experiments have demonstrated the feasibility of minimizing differential gain effects if reliable inter-calibration sources are available. POLARBEAR, for example, estimated the upper limit of these effects to be $\lesssim 0.3\%$~\cite{pb2014}, while SPTpol constrained them to be $\sim 1\%$ prior to any marginalization~\cite{Sayre:2019dic}.

Using this calibration mismatch model in Eq.~\eqref{eq:gain_drift_d}, a detector pair's TOD are
\begin{equation}
\begin{split}
d_{\t} &= \left(1+\epsilon_g\right)\left[T + Q\cos{\left(2\psi\right)} + U\sin{\left(2\psi\right)}+n\right], \\
d_{\b} &= \left(1-\epsilon_g\right)\left[T - Q\cos{\left(2\psi\right)} - U\sin{\left(2\psi\right)}+n\right],
\end{split}
\end{equation}
and the corresponding sum and difference time streams read
\begin{equation}
\begin{split}
d_{+} &= T + \epsilon_g\left[Q\cos{\left(2\psi\right)} + U\sin{\left(2\psi\right)}\right], \\
d_{-} &= \epsilon_g T + Q\cos{\left(2\psi\right)} + U\sin{\left(2\psi\right)},
\label{eq:d_minus_plus_calibration_mismatch}
\end{split}
\end{equation}
such that $T \rightarrow P$ and $P \rightarrow T$ leakage terms depend on the gain offset parameter. We expect the former leakage term to be more significant than the latter given the lower amplitude of the polarization signal, and that cross-linking during observation runs will reduce overall leakage in both $d_{+}$ and $d_{-}$.

The map-level effects of the calibration mismatch systematic for temperature and polarization are shown in Fig.~\ref{fig:syst_maps}. Expectedly, the temperature residuals are smaller compared to the polarization residuals. As this effect produces $T/P$ mixing, the temperature residual map has features similar to the spatial distribution of the polarization signal, and vice versa for the polarization maps, where the residual amplitude is $\sim 10\%$ of the input map. As with the other gain-related systematics, this effect is also expected to average-out with more frequent calibrations, more detectors, and longer observation time.

\subsection{Crosstalk}
\label{sec:crosstalk}
The last systematic effect we explore in this work is due to the experiment's electronic readout systems. Modern CMB experiments typically employ bolometric detectors operating in cryogenic environments. They adopt complex multiplexing technologies to simultaneously read out signals from many bolometers on a single readout line. This capability is required to minimize thermal losses in the cryostat that hosts the focal plane. Due to the complexity of readout technologies in cryogenic environments, the readout device can introduce a mixing of the electric signals of bolometers transported on the same readout line, an effect which is called electric \textit{crosstalk}~\cite{2012RScI...83g3113D}. We give baseline results for a readout electronic setup similar to the one employed in $\mmux$ technologies, where all 1,568 bolometers in a wafer are multiplexed together in a single SQUID (superconducting quantum interference device; used to read out the signal from the transition-edge sensors). Future experiments such as SO are expected to adopt a $\mmux$ technology and therefore have a readout scheme close to the one we simulate~\cite{Dober:2020ovt}. We also consider an alternative setup with 7 frequency-domain multiplexers ($\fmux$) with 4 SQUIDs per $\fmux$, and 28 detector pairs per SQUID. This is the reference technology for several current generation of experiment such as POLARBEAR-2/Simons Array and SPT-3G~\cite{Hattori:2015jfm,Bender:2019yna}, and has also been discussed in the context of future experiments~\cite{Crowley:2018eib}. The results we obtain with this setup are similar to those presented in the following.

The effect of crosstalk is such that the acquired raw TODs $\vd_{\bf t}$ at a given time are in reality a linear combination of the true sky measurements of each detector $\vd^{\rm det}_{\bf t}$ acquired at the same time. This can be characterized by the crosstalk leakage matrix $\mL$ as
\begin{equation}
\vd_{\bf t} = (\mathbb{1}+\mL)\vd^{\rm det}_{\bf t},
\end{equation}
where $\mathbb{1}$ is the identity matrix. To identify the signal $d_i \in \vd^{\rm det}_{\bf t}$ from each detector $i$, each detector carrier is modulated to a different readout frequency $f_i$ for detectors that are all wired together within a SQUID. A realistic representation of the element $i$, $j$ of the leakage matrix is then
\begin{equation}
L_{ij} = \frac{k_{ij}}{{\left(\Delta f_{ij}\right)}^2},
\label{eq:crosstalk_leakage_term}
\end{equation}
where $k_{ij}$ is a leakage coefficient, and the leakage depends on $\Delta f_{ij}$, the location-dependent frequency spacing between bolometers $i$ and $j$ in the focal plane~\cite{2012RScI...83g3113D}. We set $k_{ii}=0$ to avoid additional gain miscalibration, and the resulting time-stream leakage then attenuates with a constant power of 2 with respect to $\Delta f$. While there are a large number of possible modulation schemes, we use a simple linearly-spaced modulation. The readout frequencies of all bolometers within a SQUID form an arithmetic progression between a minimal and a maximal frequency $f_{\min}$ and $f_{\max}$ based on the detector's sequential placement order in rows within the SQUID. In this case, $\Delta f_{ij} \equiv \left(f_{\max}-f_{\min}\right)/n_{{\scriptstyle \mux}}$ for two consecutive detectors $i,j$ where $n_{{\scriptstyle \mux}}$ is the number of bolometers connected together within a SQUID. While this modulation model is not optimized, as bolometers which are physically near do not have the maximal possible frequency difference within the specific frequency range, the leakage amplitude proved to be dominated by the overall hardware settings (for instance resistance and induction in the readout system) such that optimizing the modulation pattern is less important. We also use $n_{\scriptstyle \mux}$ as the leakage radius so that all the bolometers within a SQUID are affected by crosstalk to achieve realistic and conservative results.

\begin{figure}[!h]
\centering
\includegraphics[width=\columnwidth]{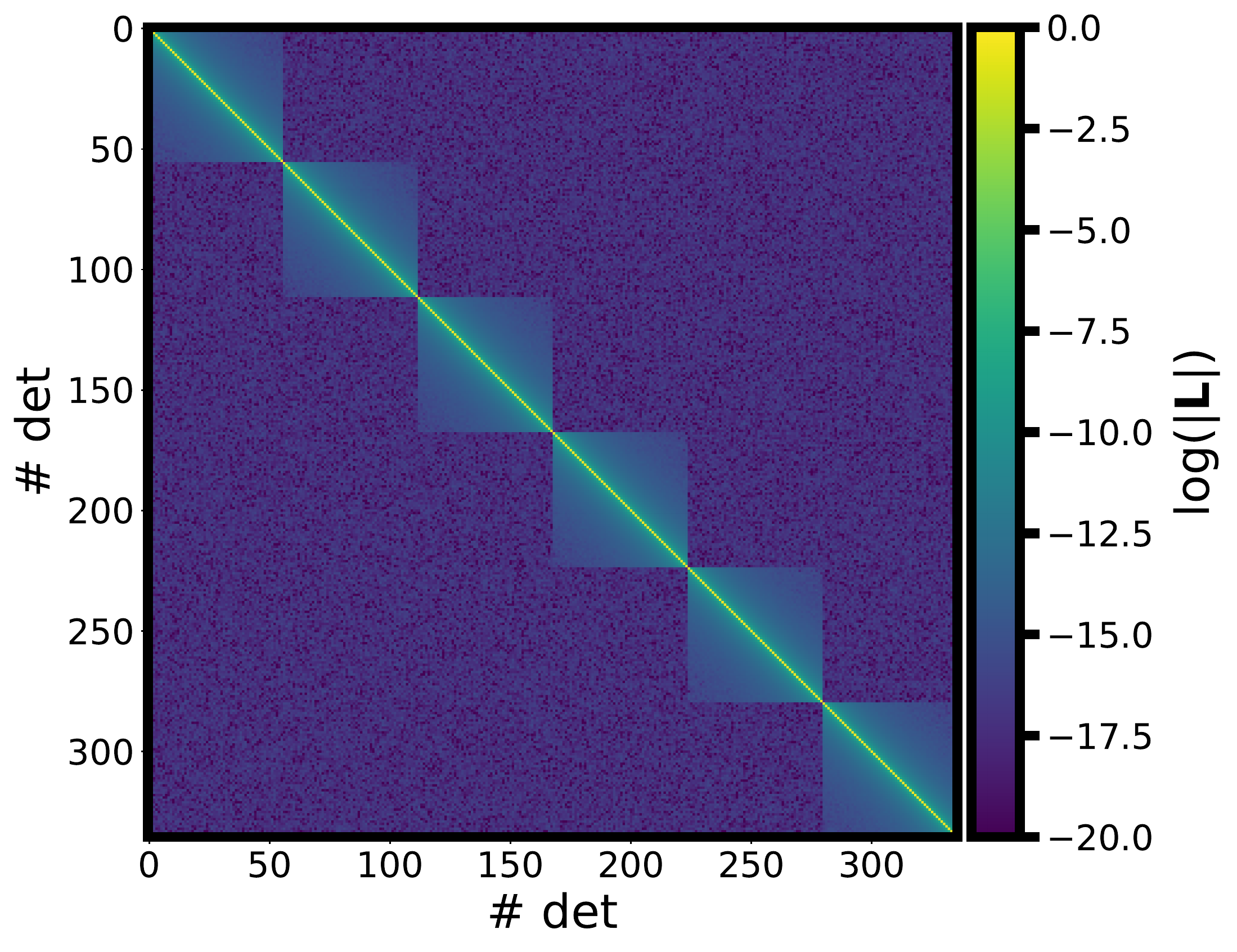}
\caption{The simulated (log) crosstalk leakage matrix $\mL$ for a subsection of the simulated detectors, given the $\fmux$ setup. The axes correspond to a detector's placement order across the focal plane. The leakage coefficients for detectors which are wired together in the same SQUID are obtained using the power-law leakage term in Eq.~\eqref{eq:crosstalk_leakage_term}. Leakage decreases as a function of frequency-distance between bolometers. The leakage appears as noise for bolometers that are sufficiently separated in frequency space. This intra-SQUID noise level is also the stochastic SQUID-to-SQUID crosstalk leakage. Diagonal elements are by default set to 1 to avoid inducing systematics that are corrected during the calibration of the bolometers. For this $\fmux$ setup, the figure shows 6 SQUIDs in a wafer, for which bolometers placed at the largest separation distance in frequency space (the top-right and bottom-left corners of each SQUID block) start having crosstalk levels which resemble the overall SQUID-to-SQUID levels. For the $\mmux$ setup, a similar plot would show more intra-SQUID correlations, as there are more detectors within a SQUID. However, due to the frequency spacing choice, most of the additional correlations would be lower than our SQUID-to-SQUID levels.}
\label{fig:crosstalk_leakage_matrix}
\end{figure}

The off-diagonal leakage coefficients $k_{ij}$ for detectors within a SQUID are drawn once for each simulation (so that the crosstalk leakage matrix remains constant during the full observation time) from a normal distribution with a $-0.03\%$ mean and a $0.01\%$ width~\cite{Crowley:2018eib}. These values are consistent with the current capabilities of the readout technologies considered for SO and CMB-S4 instruments~\cite{doi:10.1063/1.5116573,Dober:2020ovt}. The modulation frequency range is set between $f_{\min} = 4$ GHz and $f_{\max} = 8$ GHz for $\mmux$ and $f_{\min} = 1$ MHz and $f_{\max} = 5$ MHz for the $\fmux$ setup, which are typical values for these technologies. A block of the full simulated crosstalk leakage matrix $\mL$ is shown in Fig.~\ref{fig:crosstalk_leakage_matrix}. Leakages beyond the correlation radius represent the SQUID-to-SQUID crosstalk. Since this effect is subdominant, and laboratory measurements usually only provide an upper limit for it, these leakage values are drawn from a Gaussian distribution with zero mean and 0.01\% width~\cite{Crowley:2018eib}.

To understand the effect of crosstalk on the time streams, we follow the toy model of Ref.~\cite{Crowley:2018eib}. For an experiment with only two detector pairs, the (crosstalk- and noise-free) time stream $d$ of each detector is
\begin{equation}
\begin{split}
	\left.\begin{array}{l}
		d_{i} = T_1 + P_1\\[1.5ex]
		d_{j} = T_1 - P_1
	\end{array}\hspace{2mm}\right\}		& \hspace{1mm}\rotatebox[]{-90}{\text{Pair 1}}
		\\
	\left.\begin{array}{l}
		d_{k} = T_2 + P_2\\[1.5ex]
		d_{l} = T_2 - P_2
	\end{array}\hspace{2mm}\right\}		& \hspace{1mm}\rotatebox[]{-90}{\text{Pair 2}} \;\;\;,
\end{split}
\end{equation}
where we assume that both in-pair bolometers point to the same direction $\nhat_r$ so that $T_r \equiv T(\nhat_r)$ and $P_r \equiv Q(\nhat_r)\cos(2\psi_r) + U(\nhat_r)\sin(2\psi_r)$ for $r \in \{1,2\}$. The induced crosstalk leakage in each time stream is then
\begin{equation}
\begin{split}
d_{i}^{\leak} &= L_{ji}d_{j} + L_{ki}d_{k} + L_{li}d_{l} \\
d_{j}^{\leak} &= L_{ij}d_{i} + L_{kj}d_{k} + L_{lj}d_{l} \\
d_{k}^{\leak} &= L_{ik}d_{i} + L_{jk}d_{j} + L_{lk}d_{l} \\
d_{l}^{\leak} &= L_{il}d_{i} + L_{jl}d_{j} + L_{kl}d_{k},
\end{split}
\end{equation}
with no summation, such that the resulting temperature and polarization time-stream leakages are
\begin{equation}
\begin{split}
d_{+}^{\leak} =& \frac{1}{2}\left[L_{ji}+L_{ij}\right]T_1 + \frac{1}{2}\left[L_{ji}-L_{ij}\right]P_1 \\
&+ \frac{1}{2}\left[L_{ki}+L_{kj}+L_{li}+L_{lj}\right]T_2 \\
&+ \frac{1}{2}\left[L_{ki}-L_{kj}+L_{li}-L_{lj}\right]P_2, \\
d_{-}^{\leak} =& \frac{1}{2}\left[L_{ji}-L_{ij}\right]T_1 - \frac{1}{2}\left[L_{ji}+L_{ij}\right]P_1 \\
&+ \frac{1}{2}\left[L_{ki}-L_{kj}+L_{li}-L_{lj}\right]T_2 \\
&+ \frac{1}{2}\left[L_{ki}-L_{kj}-L_{li}+L_{lj}\right]P_2.
\label{eq:d_minus_plus_leak}
\end{split}
\end{equation}
A joint calibration of detectors within a pair should cancel the in-pair leakage terms, in which case $L_{ij}=L_{ji}=0$ for ${i,j}$ within a detector pair. The leakage terms then become
\begin{equation}
\begin{split}
d_{+}^{\leak} = & \frac{1}{2}\left[L_{ki}+L_{kj}+L_{li}+L_{lj}\right]T_2 \\
&+ \frac{1}{2}\left[L_{ki}-L_{kj}+L_{li}-L_{lj}\right]P_2, \\
d_{-}^{\leak} = & \frac{1}{2}\left[L_{ki}-L_{kj}+L_{li}-L_{lj}\right]T_2 \\
&+ \frac{1}{2}\left[L_{ki}-L_{kj}-L_{li}+L_{lj}\right]P_2.
\label{eq:d_minus_plus_leak_approx}
\end{split}
\end{equation}
Since $T \gg P$, the dominant temperature leakage term is $T_2 \rightarrow T_1$. The multiplicative factor of $T_2$ in this leakage term is negative, as crosstalk coefficients are mostly negative, which results in a decreased temperature power. This would also be the case if the in-pair leakage elements of Eq.~\eqref{eq:d_minus_plus_leak} are not nullified. The polarization biases are not easily estimated given the analytic leakage terms above. They depend on the specific simulated crosstalk leakage matrix and how it is coupled to the effective cross-linking with which a given sky pixel is observed. In our simulations, we keep the in-pair leakage terms for completeness, and comment on their significance in Subsec.~\ref{sec:lensing_power_biases}.

The map-level effects of the crosstalk systematic for temperature and polarization are shown in Fig.~\ref{fig:syst_maps}. Both temperature and polarization residual maps have smoothed features of the base maps, with amplitudes consistent with the induced leakage level of -0.03\%. Future experiments will employ dichroic detectors sensitive to multiple CMB frequencies at the same time, where crosstalk in the electronics will in practice generate crosstalk between the sky signal (and between its different components) at different frequencies. We did not consider this effect in this work and defer its study to future work.

Lastly, we note that other non-crosstalk-related electronic effects that are related to the readout chain may also introduce systematics that affect the low frequency part of the TOD. Detectors coupled to circuits with large time constants or data acquisition chains having a non-linear analog-to-digital (ADC) response in the electronics might distort the signal along the scan direction. It has been shown that both of these effects can be particularly complicated to deal with in the case of past experiments. Therefore, they should be given serious attention in the analysis of real data. Planck, for example, accounted for ADC non-linearities and time-constant effects in the data analysis, but showed that the major residual contamination induced by both of these effects have an important impact on the largest angular scales $\ell\lesssim 200$~\cite{planck2015-vii,planck-sroll}. These scales carry a limited weight in the lensing reconstruction, and we not to investigate them in this work. Furthermore, the typical time constants of modern detectors have a lower amplitude compared to that of Planck~\cite{arnold2012} so their impact on future CMB experiments should be less severe.

\section{Lensing analysis}
\label{sec:lensing_analysis}

We reconstruct the lensing potential of each simulation from the different ``data'' simulation sets. We perform a flat-sky quadratic estimator (QE) lensing reconstruction using the pipeline presented in Ref.~\cite{Mirmelstein:2019sxi}. After performing the standard quadratic estimator lensing reconstruction, this analysis also includes a filtering step applied to the reconstructed lensing field, which is designed to approximately minimize the corresponding power spectrum errors. The filtering is based on a patch approximation, which considers small patches within the observed area to have homogeneous noise with an effective lensing reconstruction response. This approach was shown to deliver an approximately optimal estimate of the lensing power in the presence of smoothly-varying inhomogeneous noise. We use one set of 10 systematic-free simulations to obtain an averaged reconstructed lensing power spectrum, and then repeat this calculation with sets of 10 simulations with the same CMB and noise realizations but including the effects of one of the systematics discussed in Sec.~\ref{sec:instrument_systematics_simulations}.

The lensing reconstruction stages are as follows. First, each simulation from a given set is optimally filtered using the inhomogeneous noise maps for temperature and/or polarization $\Noise$ (constructed from the pixel weights shown in Fig.~\ref{fig:weights}, with $\Noise^{-1}$ set to zero in unobserved pixels),
\begin{equation}
\begin{split}
\bar \vex &\equiv \left(\mb \cvec^{\fid} \mb^\top + \Noise\right)^{-1} \vex \\
&=\left(\cvec^{\fid}\right)^{-1}\left[\left(\cvec^{\fid}\right)^{-1}+\mb^\top \Noise^{-1}\mb\right]^{-1}\mb^\top \Noise^{-1}\vex,
\label{eq:optfilt}
\end{split}
\end{equation}
where
\begin{equation}
\vex \in \left\{
\begin{array}{cc}
T & {\rm T} \vspace{0.07in}\\
(E,B) & {\rm P} \vspace{0.07in}\\
(T,E,B) & {\rm MV}
\end{array}
\right.
\end{equation}
is a vector of the CMB maps, $\mb$ is the transfer function (a CS Gaussian beam with $\sigma_{{\scriptstyle \FWHM}}$ width) and $\cvec^{\fid}$ is a set of fiducial lensed power spectra which were obtained from \CAMB\footnote{\href{https://camb.info/}{https://camb.info/}.}~\cite{Lewis:1999bs}. Eq.~\eqref{eq:optfilt} is solved using the multi-grid-preconditioned conjugate gradient method~\cite{Smith:2007rg,Ade:2013tyw,Story:2014hni,Ade:2015zua,PL2018}. Then, the filtered simulations from each ``data'' set are used to estimate $\hat{\phi}(\vecx)$, the unnormalized QE, using
\begin{equation}
\begin{split}
\hat{\phi}(\vecx) = \frac{1}{2}{\bar\vex}^\top \frac{\delta \mC^{\vex\vex}}{\delta \phi(\vecx)}\bar\vex ,
\end{split}
\end{equation}
where $\mC^{\vex\vex}$ is the covariance of the map $\vex$~\cite{Hanson:2009kr}. This QE is biased by non-zero average values of statistical anisotropy in the map (due to e.g. sky masking and noise anisotropy). This mean field (MF) bias, $\langle\hat{\phi}\rangle_{\MC}$, is subtracted from the lensing estimator $\hat{\phi}$. The unbiased, and unnormalized, QE is then converted to the convergence estimator
\begin{equation}
\hat{\kappa}_{\vecL} \equiv \left(\hat{\phi}_{\vecL}-{\langle\hat{\phi}_{\vecL}\rangle}_{\MC}\right) \times \frac{2}{L(L+1)}.
\end{equation}
This is then filtered using the effective patch-approximated response $\boldsymbol{\R}_{\eff}^{\kappa}$, the reconstruction noise $\mN^\kappa_{0,\eff}$ (see Ref.~\cite{Mirmelstein:2019sxi}) and a fiducial $\kappa$ spectrum $\mC^{\kappa\kappa}_{\fid}$, to define
\begin{equation}
\hat{\boldsymbol{\kappa}}^{\filt} \equiv \mC^{\kappa\kappa}_{\fid}\left(\mC^{\kappa\kappa}_{\fid}+\mN^\kappa_{0,\eff}\right)^{-1} \left(\boldsymbol{\R}_{\eff}^{\kappa} \right)^{-1} \hat{\boldsymbol{\kappa}}.
\label{eq:kappa_filt}
\end{equation}
The QE is also normalized in this step using the effective response. This additional filtering is specifically performed on the convergence ($\kappa$) map and not directly on $\hat\phi$ as the $\kappa$ reconstruction is approximately local in real space and has approximately white noise. The (noise biased) lensing power spectrum is then obtained from the filtered $\kappa$ maps,
\beqa
C_L^{\hat{\phi}_1\hat{\phi}_2} &\equiv& \frac{4}{\fpatch n_L L^2(L+1)^2}\sum\limits_{\vecell {\text{ in $\vecL$ bin}}}\hat{\kappa}_{1,\vecell}^{\filt}\left({{\hat{\kappa}_{2,\vecell}}^{\filt}}\right)^{*}, \nonumber\\
\label{eq:unbiased_estimator}
\enqa
where $n_L$ is the number of modes on the flat sky assigned to lensing multipole $L$ in our simulation maps\footnote{In a full-sky analysis, $n_L = 2L+1$.} and
\beqa
\fpatch = \sum_{p} f_{p} \left(\frac{\R_L^{\kappa,p}}{\R_L^{\kappa,\fid}}\right)^2
\label{eq:f_A}
\enqa
is the required normalization for our analytic patch approximation estimator~\cite{Mirmelstein:2019sxi}. $f_p$ is the fraction of the map area in patch $p$. The MF is calculated twice, from two sets of 48 MC simulations. The subscripts of $\kappa$ and $\phi$ in Eq.~\eqref{eq:unbiased_estimator} indicate the MF set which was used to debias each estimator. Each MF estimate has independent MC noise, so the lensing power spectrum calculated from a pair of MF-subtracted QEs has no MC noise biases. We do not include systematics in the MF simulations during the analysis, but we comment on this possibility in Sec.~\ref{sec:mitigation_techniques}.

Since the lensing power spectrum estimator is a 4-point correlation function, it has a disconnected bias arising from the correlation of Gaussian fields, $N_{0,L}^{\phi\phi}$. There is an additional bias term, $N_{1,L}^{\phi\phi}$, resulting from connected contractions that are not proportional to the lensing spectrum at $L$. Both terms can be modeled analytically to correct the obtained lensing power spectrum~\cite{Kesden:2003cc,2011PhRvD..83d3005H}, although calculating a realization-dependent $N_{0,L}^{\phi\phi}$ and a $N_{1,L}^{\phi\phi}$ term using the patch approximation corrects the reconstruction biases more optimally. We subtract an estimate of $N_{0,L}^{\phi\phi}$ from each power spectrum estimate, with the respective realization-dependent estimate $\RDN_{0,L}^{\phi\phi}$ obtained from a set of 480 MC simulations~\cite{Story:2014hni,Ade:2015zua},
\beqa
\RDN_{0,L}^{\hat\phi\hat\phi} &=&\nonumber\\
\Bigl\langle &-&C_{L}^{\hat{\phi}\hat{\phi}}\left[\bar{\vex}_{\MC_{1}^{\phi_1}},\bar{\vex}_{\MC_{2}^{\phi_2}},\bar{\vex}_{\MC_{2}^{\phi_2}},\bar{\vex}_{\MC_{1}^{\phi_1}}\right] \nonumber\\
&-&C_{L}^{\hat{\phi}\hat{\phi}}\left[\bar{\vex}_{\MC_{1}^{\phi_1}},\bar{\vex}_{\MC_{2}^{\phi_2}},\bar{\vex}_{\MC_{1}^{\phi_1}},\bar{\vex}_{\MC_{2}^{\phi_2}}\right] \nonumber\\
&+&C_{L}^{\hat{\phi}\hat{\phi}}\left[\bar{\vex}_{\MC_{1}^{\phi_1}},\bar{\vex}_{\dat},\bar{\vex}_{\dat},\bar{\vex}_{\MC_{1}^{\phi_1}}\right]\nonumber\\
&+&C_{L}^{\hat{\phi}\hat{\phi}}\left[\bar{\vex}_{\dat},\bar{\vex}_{\MC_{1}^{\phi_1}},\bar{\vex}_{\MC_{1}^{\phi_1}},\bar{\vex}_{\dat}\right] \nonumber\\
&+&C_{L}^{\hat{\phi}\hat{\phi}}\left[\bar{\vex}_{\dat},\bar{\vex}_{\MC_{1}^{\phi_1}},\bar{\vex}_{\dat},\bar{\vex}_{\MC_{1}^{\phi_1}}\right]\nonumber\\
&+&C_{L}^{\hat{\phi}\hat{\phi}}\left[\bar{\vex}_{\MC_{1}^{\phi_1}},\bar{\vex}_{\dat},\bar{\vex}_{\MC_{1}^{\phi_1}},\bar{\vex}_{\dat}\right] \Bigr\rangle_{\MC_{1}^{\phi_1},\MC_{2}^{\phi_2}}\nonumber\\
\enqa
where $\bar{\vex}_{\dat}$ is the vector of our ``data'' simulations and the 1 and 2 subscripts refer to the matching CMB and lensing potential realizations of a given MC simulation. Using the realization-dependent debiasing term rather than a general MC $N_{0,L}^{\phi\phi}$ is crucial, as it automatically mitigates systematic biases that arise entirely from small changes to the noise and CMB power spectra that enter the disconnected bias. We show the differences between $\RDN_{0,L}^{\phi\phi}$ from ``data'' simulations with and without systematics in Sec.~\ref{sec:lensing_power_biases}.

The debiased lensing power is then
\beqa
\hat{C}_L^{\phi\phi} &\equiv& C_L^{\hat{\phi}_1\hat{\phi}_2} - \RDN_{0,L}^{\hat\phi\hat\phi}.
\label{eq:lensing_power}
\enqa
We do not debias the reconstructed power using $\MCN_{1,L}^{\phi\phi}$ as this term would vanish when differencing power spectra with and without systematics (because our MC simulations are systematics-free).

Lastly, all 10 power spectra from each set are averaged and compared to the averaged systematics-free power spectrum to assess how the systematics affect the reconstructed CMB lensing power spectrum. We perform temperature-only (T), polarization-only (P) and minimum-variance (MV) temperature+polarization lensing reconstructions for each ``data'' set to show how the systematics bias different estimators. The resulting systematic effects on lensing reconstruction are shown and discussed in the following section.

\section{Systematics biases}
\label{sec:systematics_biases}

\subsection{CMB power spectrum biases}
\label{sec:map_power_biases}

We start by examining the effects of the systematics described in Sec.~\ref{sec:instrument_systematics_simulations} on the temperature and polarization power spectra $C_\ell$. Apart from providing a good consistency check, these power spectra can help us understand the nature of some of the induced lensing biases and hence suggest possible mitigation techniques.

\begin{figure*}[!ht]
\includegraphics[width=.9\textwidth]{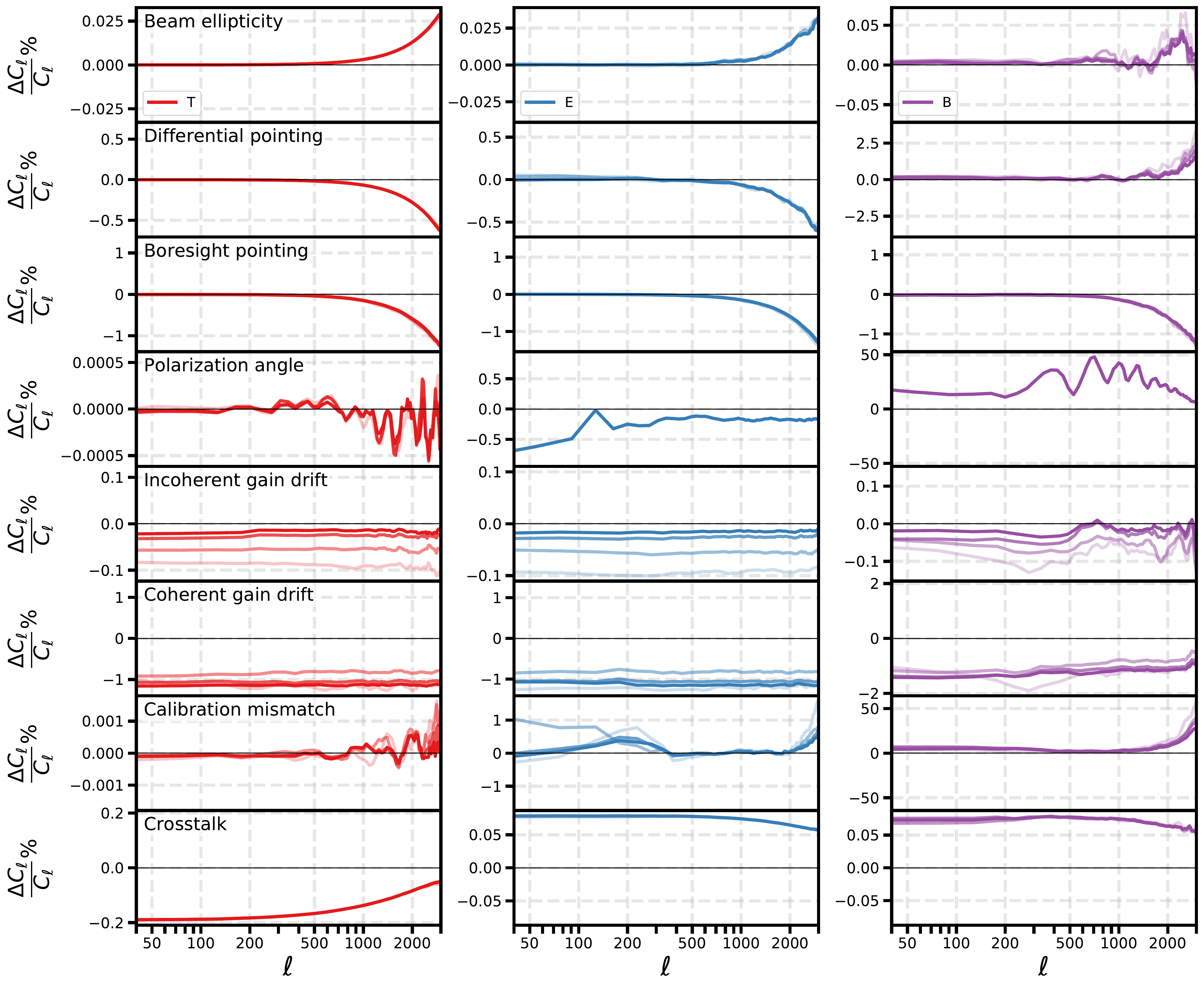}
\caption{The fractional differences between the $T$ (red lines), $E$ (blue lines) and $B$ (purple lines) power spectra $C_\ell$ with and without systematics in the multipole range $40<\ell<3000$. The bright to dark curve shades correspond to observation times of 3, 6, 9 and 12 days respectively. Curves were smoothed with $\sigma_{\scriptscriptstyle \ell}=5$ to highlight the bias differences for different observation times. For temperature, $C_\ell$ is mostly signal-dominated, while for the $B$-mode it is noise-dominated. Noise becomes dominant for the $E$-mode spectrum at $\ell \gtrsim 2000$.}
\label{fig:C_ell_nosyst_vs_C_ell}
\end{figure*}

We first calculate the pseudo power spectra $\pC_\ell$s of the flat-sky maps using the discrete 2D Fourier components of the weighted temperature or polarization maps, $a_{\scriptscriptstyle \vecell}$,
\begin{equation}
\pC_\ell \equiv \frac{1}{n_{\scriptscriptstyle \ell} b_{\scriptscriptstyle \ell}^2} \sum\limits_{\ell~{\rm in}~\vecell~{\rm bin}} a_{\scriptscriptstyle \vecell}a_{\scriptscriptstyle \vecell}^{*},
\end{equation}
where $n_{\scriptscriptstyle \ell}$ is the number of modes on the flat sky assigned to the multipole $\ell$ in our simulation maps. To obtain an unbiased estimate of $C_\ell$ we then deconvolve the effect of the sky mask using the MASTER approach, using a pure estimator to avoid $E/B$ mixing~\cite{Smith:2005gi} in the polarization field as implemented in the publicly available code \namaster\footnote{\url{https://github.com/LSSTDESC/NaMaster}}~\cite{master,namaster} (for the lensing reconstruction, our first optimal filtering step optimally suppresses $E/B$ mixing variance as the filter includes the full noise and mask inhomogeneity, so no further $E/B$ projection is required). The fractional differences between $C_\ell$s of maps with and without systematics are shown in Fig.~\ref{fig:C_ell_nosyst_vs_C_ell}. These power spectra were computed from noise-free simulations in order to highlight the impact of the systematics on the signal.

The fractional differences in the power spectra due to beam ellipticity, differential pointing, and boresight pointing systematics have a similar shape to a beam transfer function, especially for the temperature and $E$-mode spectra. The $B$-mode spectrum residual shapes for beam ellipticity and differential pointing are affected by leakage from $T$ and $E$, which are large relative to the $B$-mode power amplitude, and causes them to have a somewhat different shape.
In practice, the effects of these systematics, combined with the scanning strategy, produce a modified smoothing to the map which is not corrected by the CS beam transfer function $b_{\scriptscriptstyle \ell}$ used for constructing the power spectra (in our beam-related systematic analyses, the reference beam window function does not account for these beam-like effects, nor do we include any beam uncertainties in the analysis). For the beam ellipticity and differential pointing systematics, this bias stems from the leakage terms of Eq.~\eqref{eq:d_minus_plus_explicit} which are coupled to $b_{+}$. The increase of power observed at small scales induced by the beam ellipticity systematic is consistent with the fact that an elliptical beam, whose axes are given by Eq.~\eqref{eq:perturbed_beam}, has effectively a smaller average width than a circular Gaussian beam of width $\sigma_{\CS}$ (the average width of the elliptical beam is taken as $\sqrt{\sigma_{\min}\sigma_{\maj}}$).

The boresight pointing systematic smoothing stems from the nature of the systematic itself: jitters during an observation run induce additional smoothing in the map. Fig.~\ref{fig:C_ell_nosyst_vs_C_ell} also shows the fractional differences for 3, 6, 9 and 12 days of observation. For these three systematics, the biases remain relatively constant in time and do not average out. The beam ellipticity systematic produces the smallest biases compared to the other systematics we simulate. The lensing bias induced by these effective beam mismatches is largely corrected at the lensing reconstruction step by $\RDN_{0,L}$, as we show in Sec.~\ref{sec:mitigation_techniques}. Using a beam window function with an effective width tailored to each of these systematics in the lensing reconstructing analysis should also mitigate most of their biases that originate from differences in power (see Sec.~\ref{sec:mitigation_techniques}).

The polarization angle biases are quite substantial for the polarization power spectra. These biases are characterized well by the analytic approximations~\cite{2013ApJ...762L..23K} for an effective constant angle perturbation $\Delta\psi$,
\begin{equation}
\begin{split}
C_\ell^{\tilde{E}\tilde{E}} &= \cos^2\left(2\Delta\psi\right)C_\ell^{EE}-\sin^2\left(2\Delta\psi\right)C_\ell^{BB}, \\
C_\ell^{\tilde{B}\tilde{B}} &= \sin^2\left(2\Delta\psi\right)C_\ell^{EE}+\cos^2\left(2\Delta\psi\right)C_\ell^{BB},
\end{split}
\end{equation}
where $\tilde{E}$ and $\tilde{B}$ are the perturbed polarization modes. From these equations, we see that the large $C_\ell^{BB}$ bias is mostly the $C_\ell^{EE}$ power spectrum, scaled by a constant which depends on an effective polarization angle error, while the $C_\ell^{EE}$ bias is an effective gain which also depends on this error, as the $B \rightarrow E$ leakage term is sub-dominant. As the analytic approximations describe these biases well for an effective $\Delta\psi$ despite each detector having a different polarization angle error, they may be used to sufficiently mitigate these biases (see Sec.~\ref{sec:mitigation_techniques} for more details on this bias mitigation).

The incoherent gain drift biases evidently decrease with increased observation time. The power spectra from the full 12 observation days have a negligible bias relative to the $C_\ell$ amplitudes for temperature and polarization. For the coherent gain drift, however, more frequent gain calibrations are required for the biases to average out, or a longer observation time that improves the overall cross-linking. While a long-lasting coherent gain drift induces relatively significant biases, the majority of this effect would be identified and mitigated during early stages of an experiment's data analysis prior to the lensing reconstruction. For example, it is possible to correct for this bias during the map-making stage using the signal variations of bolometers inside the cryostat that are not coupled to the optical chain, as those are insensitive to the sky signal. Our estimates for this systematic effect are therefore pessimistic. Both of these gain systematics have a relatively constant amplitude effect on the spectra across the considered $\ell$-range, as expected from the mean gain described in Subsec.~\ref{sec:gain_drifts}.

For the calibration mismatch systematic biases, while these also decrease with longer observation times, they continue to be significant for the polarization spectra after 12 observation days. Expectedly, the temperature power spectrum biases are small, as they stem from $P \rightarrow T$ leakage which is small compared to the temperature power amplitude. The $T \rightarrow P$ leakage, however, is quite substantial, especially for the mid-$\ell$-range $B$-mode power spectrum. As the $B$-mode spectrum is noise-dominated at $\ell>1000$, the large bias in that multipole range is not very significant.

As seen in Fig.~\ref{fig:C_ell_nosyst_vs_C_ell}, crosstalk is the only systematic for which the temperature biases are higher than the polarization biases. Excluding the in-pair leakage terms of Eq.~\eqref{eq:d_minus_plus_leak} in both temperature and polarization time streams results in a $\sim30\%$ bias decrease. Since the overall leakage is already quite negligible, we reconstruct the lensing potential from systematics which include the in-pair leakage terms. As discussed in Subsec.~\ref{sec:crosstalk}, the temperature power spectrum bias is negative. For our specifications the resulting polarization power spectra have additional power.

\subsection{Lensing power biases}
\label{sec:lensing_power_biases}

\begin{figure}[!h]
\centering
\includegraphics[width=\columnwidth]{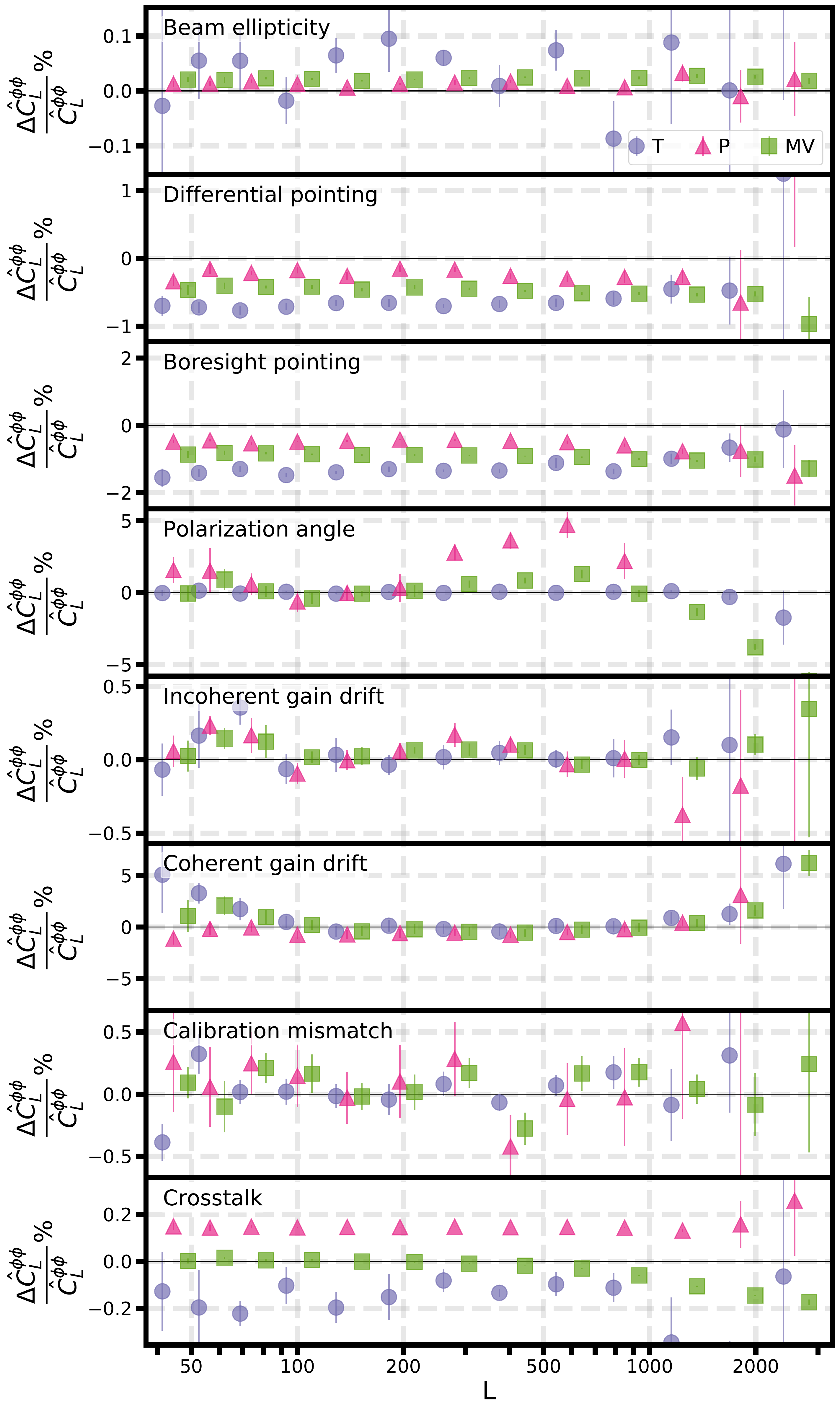}
\caption{The fractional differences between $\hat{C}_L^{\phi\phi}$ with and without systematics for T (purple circles), P (pink triangles), and MV (green squares) lensing reconstructions. The reconstruction noise dominates at $L\gtrsim200$ for T and P reconstructions, and at $L\gtrsim300$ for MV reconstruction. The 13 bin widths are log-spaced between $10$ and $1000$.}
\label{fig:C_L_phiphi_nosyst_vs_C_L_phiphi}
\end{figure}

As there are various ways in which the lensing power can be used for constraining cosmological observables, it is useful to show the significance of the systematics-induced lensing biases in several ways. We first demonstrate how significant these biases are with respect to the lensing power spectrum. The fractional differences between the averaged reconstructed lensing power from simulations with and without systematics for T, P and MV reconstructions are shown in Fig.~\ref{fig:C_L_phiphi_nosyst_vs_C_L_phiphi}.

All bias amplitudes are below the 5\% level compared to the lensing power, with most under 0.5\%. These levels are generally consistent with their expected values from the $C_\ell$-level biases. The main difference between the $C_\ell$ and the $\hat{C}_L^{\phi\phi}$ biases is in their shapes. Beam-like $C_\ell$ biases appear as a bias on the $\phi$ power spectrum amplitude. This is expected to be roughly constant on large scales, as the $\hat{\phi}$ estimator is normalized using biased fiducial CMB power spectra, so the resulting $\hat{C}_L^{\phi\phi}$ have a different amplitude. Moreover, for beam-like systematics the MV reconstruction biases appear to be bounded by the T and P biases. The most significant bias of these cases comes from the boresight pointing systematic, which is at a 1\% level for an MV reconstruction. As with the power spectra, the beam ellipticity biases on the reconstructed lensing power spectra are negligible.

The polarization-only lensing reconstruction is most problematic in the presence of unmitigated polarization angle errors. From our pessimistic probability distribution for $\Delta\psi$, the resulting amplitude of polarization-only lensing biases is up to $\sim 5\%$ for $L<1000$. MV reconstruction benefits from the low $P\rightarrow T$ leakage, and the bias levels remain below $\sim 1.5\%$ for the same multipole range.

The gain-related systematics biases are randomly scattered around zero with varying levels of significance. As with the $C_\ell$, the most prominent bias is that of the coherent gain drift. Its temperature-only reconstructed power is $\sim5\%$ higher compared to the systematics-free reconstructed power in the signal-dominant multipole range, although with a large uncertainty. Its MV reconstruction bias is $\lesssim 2.5\%$.

For crosstalk, while the reconstruction biases are consistently below 0.3\%, the MV reconstruction proves to be the least biased over the signal-dominated $L$-range. This is most likely due to the opposite signs of the biases in T and P reconstructions, which seems to cancel in the combined reconstruction.

The only systematics for which the MV biases are the smallest of the three are incoherent gain drift, calibration mismatch and crosstalk.

Another way to quantify systematic-induced lensing biases is by performing a likelihood analysis to estimate their detectability in the lensing spectrum. Although significance values estimated from our 12-day scaled-noise simulations are not expected to correspond to what an experiment with a realistic observing time would see, they provide a useful reference point. We use the simplified log-likelihood
\begin{equation}
\ln \mathcal{L} = -\sum\limits_{\lb}\frac{A^2\left(\hat{C}_{\lb}^{\phi\phi,\syst}-\hat{C}_{\lb}^{\phi\phi}\right)^2}{2\sigma_{\hat{C}_{\lb}^{\phi\phi}}^2},
\end{equation}
where $\hat{C}_{\lb}^{\phi\phi,\syst}$ and $\hat{C}_{\lb}^{\phi\phi}$ are the reconstructed lensing power spectra with and without systematic effects, respectively, in a specific multipole bin $\lb$. The parameter $A$ is the amplitude parameter for the bias with uncertainty $\sigma_A$, which quantifies how significant the bias is compared to the reconstructed lensing power error bar $\sigma_{\hat{C}_{\lb}^{\phi\phi}}$. The second derivative of $\mathcal{L}$ with respect to $A$ is the inverse variance of $A$, $\sigma_A^{-2}$, such that
\begin{equation}
\sigma_A = \left[\sum\limits_{\lb}\frac{\left(\hat{C}_{\lb}^{\phi\phi,\syst}-\hat{C}_{\lb}^{\phi\phi}\right)^2}{\sigma_{\hat{C}_{\lb}^{\phi\phi}}^2} \right]^{-\frac{1}{2}}.
\label{eq:sigma_A}
\end{equation}
A constant bias for which $\sigma_A<1$ will be detectable by more than $1\sigma$, and vice versa. The values of $\sigma_A^{-1}$ for the different systematics are shown in Table~\ref{table:sigma_A}.
For the study-case we considered, the only systematic that can be detected by more than $1\sigma$ is the polarization angle systematic. For this systematic, the unmitigated polarization-only bias detection level is the highest, however including the temperature map in the analysis significantly reduces the bias significance. Only the boresight pointing, coherent gain drift and polarization angle systematics produce biases with detection levels above $0.5\sigma$. Unlike the polarization angle systematic, for coherent gain drift the highest bias detection level occurs when using the temperature map alone to reconstruct the lensing potential, while for polarization angle this is the case for only for the polarization-only reconstruction.

For differential beam ellipticity, we found that the leakage reduced to significantly below the detection level, mainly due to the number of bolometers used. An experiment with the same beam width and only 10-100 detectors, or larger beam-width and similar number of detectors, would be more affected by this systematic.

Apart from the polarization angle systematic, the biases resulting from our coherent gain drift model also seem to be relatively problematic for SO and future CMB experiments. This is not the case for the incoherent drift, mainly because scanning the sky repeatedly with a large number of detectors, each of which has a different gain drift, helps to average out the effect. Performing gain calibrations at shorter time intervals, or observing each sky area more times, may mitigate some of the effect of coherent drifts. On the other hand, experiments using a scanning strategy with less cross-linking may find a larger effect.

\begin{table}[ht]\vspace{0.2in}
\begin{center}
\begin{tabular}{ l ccc }
\hspace{0.02in}			Systematics														&	T 							&	P 					 	&	MV 	\EndTableHeader\\
			\hline
			\hline
\hspace{0.02in}			Beam ellipticity			\hspace{0.46in}					&	\ApplyGradient{0.06}	&	\ApplyGradient{0.00}	&	\ApplyGradient{0.01}	\\
\hspace{0.02in}			Differential pointing 											&	\ApplyGradient{0.27}	&	\ApplyGradient{0.09}	&	\ApplyGradient{0.28}	\\
\hspace{0.02in}			Boresight pointing												&	\ApplyGradient{0.52}	&	\ApplyGradient{0.20}	&	\ApplyGradient{0.52}	\\
\hspace{0.02in}			Polarization angle												&	\ApplyGradient{0.05}	&	\ApplyGradient{2.20}	&	\ApplyGradient{0.60}	\\
\hspace{0.02in}			Incoherent gain drift											&	\ApplyGradient{0.36}	&	\ApplyGradient{0.05}	&	\ApplyGradient{0.04}	\\
\hspace{0.02in}			Coherent gain drift												&	\ApplyGradient{0.56}	&	\ApplyGradient{0.28}	&	\ApplyGradient{0.64}	\\
\hspace{0.02in}			Calibration mismatch											&	\ApplyGradient{0.38}	&	\ApplyGradient{0.11}	&	\ApplyGradient{0.09}	\\ 
\hspace{0.02in}			Crosstalk															&	\ApplyGradient{0.11}	&	\ApplyGradient{0.06}	&	\ApplyGradient{0.03}	\\ 
\end{tabular}
\caption{Detection significance of systematics biases with respect to the lensing power uncertainty for T, P and MV reconstructions. The values in the table are calculated using Eq.~\eqref{eq:sigma_A}. The values are color-coded from most significant biases (darker red) to less significant (lighter red). The detection significance for the polarization-only reconstruction beam ellipticity bias is $\sim1 \times 10^{-3}$. Assuming that all the biases are independent, the combined bias is measured with a $\sim0.9\sigma$ significance for MV and T reconstructions.}
\label{table:sigma_A}
\end{center}
\end{table}

Our chosen parameters for modeling the calibration mismatch are relatively pessimistic, as most CMB experiment have a lower gain uncertainty. While this systematic can potentially be a problem, it is evident that for our specifications, especially the number of detectors and scanning strategy, even this pessimistic case does not affect our reconstructed lensing power spectra in an important way. Decreasing the gain uncertainties by a factor of 10 to $\approx 1\%$ compared to the baseline case shown here would lead to $C_L^{\phi\phi}$-level biases lower than $\sim0.1\%$ for all three reconstruction setups and detection levels of 0.18 (T), 0.01 (P), and 0.03 (MV). A moderate improvement compared to the pessimistic case we assumed should thus already be sufficient to mitigate this bias to an acceptable level, although a more realistic scanning time may also be sufficient.

Our crosstalk simulations use realistic but relatively pessimistic parameters. Excluding in-pair leakage terms, which can be usually corrected when performing in-pair gain calibrations, reduces the $C_{\ell}$ bias levels by 30\%. Optimizing the frequency spacing of the different bolometers can also be achieved to establish lower leakage levels. Since the overall bias levels we show are very low, we did not perform further optimization to our crosstalk modeling. Increasing the crosstalk leakage coefficient distribution's mean and width by a factor of 10, the $C_{\ell}$ bias levels also increased by a factor of 10, although the significance stays below $\sim 1\sigma$. This suggests that controlling the crosstalk leakage levels to about $-0.3\%$ is sufficient for the purpose of lensing reconstruction. This crosstalk level is higher than the expected performance of future-generation instruments based on $\mmux$ technologies. We therefore conclude that crosstalk is not expected to become a major systematic for lensing. Moreover, since crosstalk is constant in time (as it mainly depends on the wiring of the electronics), it should be possible to account for its potential biases, at least partly, in the simulations used to evaluate the mean field of the quadratic estimator if needed (see Sec.~\ref{sec:mitigation_techniques} for more details). For the most extreme scenarios, the crosstalk leakage matrix can be estimated from dedicated calibration data and used to correct for its effect at the time-stream level prior to the map-making step, but at the cost of inducing correlated noise~\cite{Henning:2017nuy}.

\begin{figure}[!htbp]
\centering
\includegraphics[width=\columnwidth]{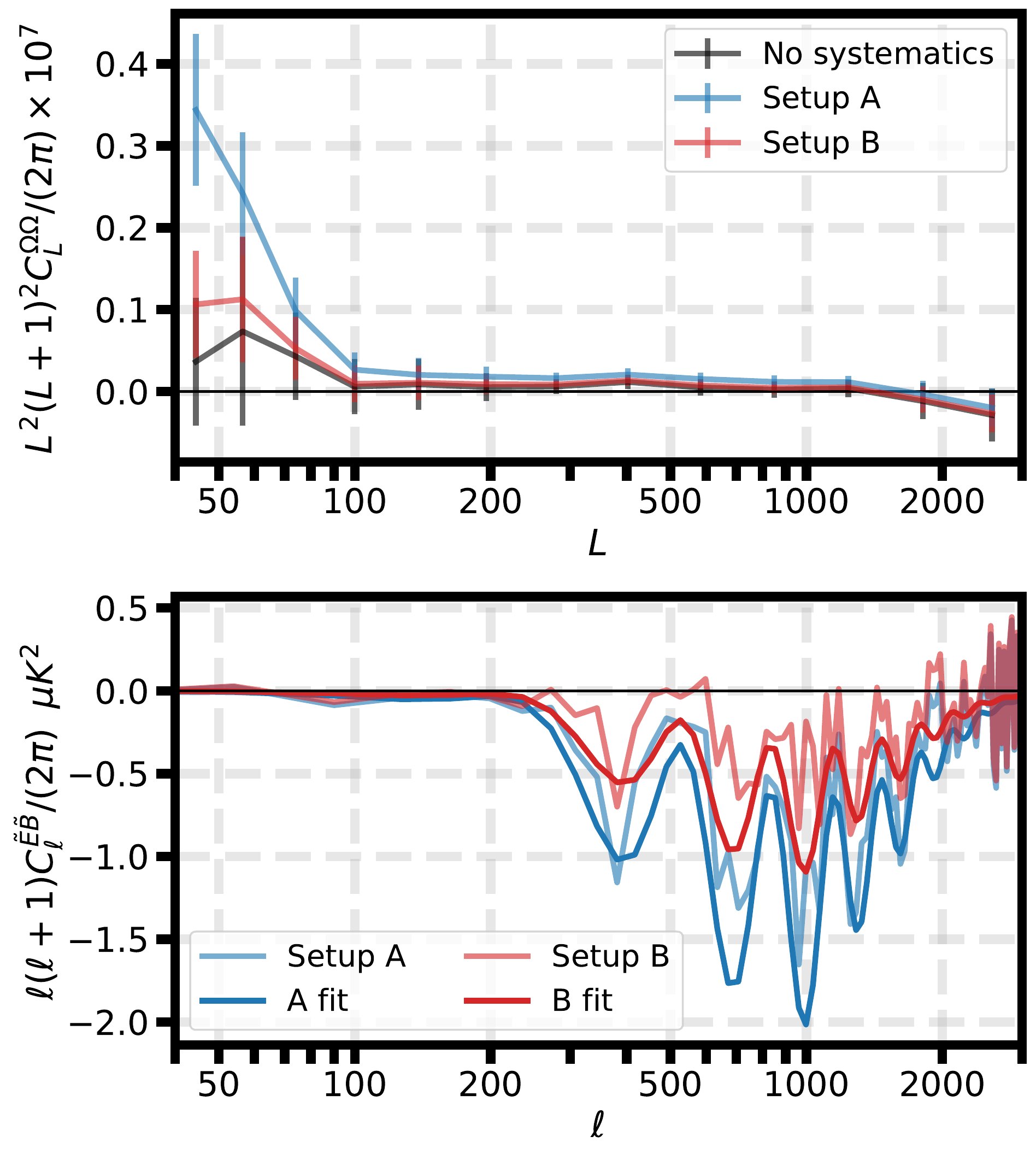}
\caption{
\textit{Top}: The lensing curl signal induced by the polarization angle systematic for a polarization-only reconstruction. For setup A ($\Delta\psi\in \N(-1.1\degree,0.5\degree)$, blue line), this signal is detectable by just over $2\sigma$ at $L\lesssim 50$. For setup B ($\Delta\psi\in \N(-0.5\degree,2.0\degree)$, red line), this signal has $\lesssim 1\sigma$ detection level for $L>40$. The polarization angle shift of setup A was used as a baseline in the lensing biases analysis.
\textit{Bottom}: $C_\ell^{\tilde{E}\tilde{B}}$ power spectra of one simulation induced by a miscalibration of the polarization angles of the detectors for setup A (blue) and setup B (red). The darker solid lines for each power spectrum show the power spectra corresponding to the best fit value of $\Delta\psi$ obtained by fitting the analytic approximation of Eq.~\eqref{eq:pol_ang_powers} to the simulated $C_\ell^{\tilde{E}\tilde{B}}$ and assuming the theoretical $E$ and $B$ power spectra expected from the underlying cosmological model. The recovered effective values of $\Delta\psi$ for each fit deviate from the mean of the input error by up to $\sim 6\%$.
}
\label{fig:polang_eb_curl}
\end{figure}

The last thing we consider for characterizing systematics-induced lensing biases relates to the lensing curl signal. The CMB photon deflection field $\mathbf{d}$ is a vector field defined on the sphere, and as such it can be written as a combination of a gradient and a curl-like mode, $\mathbf{d}=\boldsymbol\nabla\phi+\boldsymbol\star\boldsymbol\nabla\Omega$, where $\phi$ is the lensing potential and $\Omega$ the curl potential\footnote{We recall that in two dimensions and in the flat sky approximation $\boldsymbol\star\hat{e}_x=\hat{e}_y$ and $\boldsymbol\star\hat{e}_y=-\hat{e}_x$.}. In addition to biases in the lensing potential power spectrum, we also tested whether instrumental systematics produce a non-zero lensing curl signal. While cosmological curl signal is already expected to be non-zero from second-order lensing effects~\cite{Hirata:2003ka,Cooray:2002mj,Pratten:2016dsm,Fabbian:2017wfp}, these would remain undetectable in the curl power spectrum for the foreseeable future. As for the lensing potential, the lensing curl mode can also be reconstructed using the quadratic estimators~\cite{Namikawa:2011cs}. We use a pipeline analogous to the one described in Sec.~\ref{sec:lensing_analysis} for $\phi$, but using the lensing response functions relevant to $\Omega$. We found that the only systematic which produces a significant non-zero lensing curl signal is the polarization angle miscalibration. The resulting curl signal, shown in Fig.~\ref{fig:polang_eb_curl}, manifests more significantly at large scales, for $L\lesssim50$. When the polarization angle perturbations are drawn from the less pessimistic distribution of setup B, this signal is less prominent and remains below detection levels. This suggests that having a non-zero curl signal could be a useful tool for diagnosing problems with the calibration of polarization angles.

\section{Mitigation techniques}
\label{sec:mitigation_techniques}

Systematics mitigation can generally be performed at different levels, from instrument planning through data collection to the final analysis stage. In this section, we focus on mitigation techniques performed at the analysis stage. This is mainly motivated by the results of our work, which demonstrate that our realistic and conservative assumptions on instrument specifications already yield relatively small bias levels. 

Before discussing analysis-level mitigation techniques, we first briefly discuss how an experiment's scanning strategy affects systematic biases. Some systematic biases are automatically mitigated by scanning the same region of the sky from different directions. Each time a given sky pixel is observed by a different detector pair, the final map value in the pixel is less sensitive to systematic variations between detectors (as well as a reduced instrument white-noise level). Observing the same sky area with the same detector pair also contributes toward mitigation, as a given detector pair may also have systematics that vary randomly in time. This is important when an experiment plans its scanning strategy, as there is a trade-off between repeated observation of specific areas in a given time frame, and using the same given time frame to scan more areas of the sky at the expense of reduced cross-linking.

In our simulations, we modeled observations over a relatively small sky patch, within which most CESs had some overlapping region. For the systematics that do not depend on properties of the instrument that are constant in time, we found that the process of repeated observation over the same area reduces most of the biases in the CMB maps and power spectra, which in turn also reduces the biases on the lensing reconstruction power. A scanning strategy can also be devised to mitigate specific systematic biases. For example, differential pointing and differential gain systematics can be mitigated by introducing a boresight rotation to the scanning strategy~\cite{Thomas:2019hak}. It may also be possible to mitigate differential pointing effects by knowing analytically how the scanning strategy couples to the pointing signal~\cite{McCallum:2020jsp}. To avoid experiment-specific conclusions, in this work we adopted the most conservative approach and did not try to implement scans that are optimized to mitigate systematic effects (as done for several instruments); most of the biases we find are in any case only of marginal importance. Below, we also show that the differential pointing and polarization angle biases can be mitigated also at the analysis stage. Mitigation at the analysis stage may be more generally applicable, as it does not depend strongly on a given experiment's scanning strategy or other specifications.

As for mitigating systematic biases at the analysis stage, we first discuss techniques that are potentially helpful at the map-making level. Solving Eq.~\eqref{eq:map_making} can be done more optimally by including deprojection terms in the signal vector, or by employing filters to mitigate unwanted signal contaminants. In our efforts to mitigate some of the gain drift biases, we tried using a simple deprojection technique to solve for an additional gain which contaminates the polarization maps~\cite{Ade:2014afa}, but depends only on the sky pointing. This is done by solving for an additional sky component $G$ which enters the polarization time stream as
\begin{equation}
d_{-} = G + Q\cos\left(2\psi\right) + U\sin\left(2\psi\right).
\end{equation}
Our simulated gain variation is not constant in time, so this deprojection model did not mitigate any of the biases we observe. Using different deprojection methods, such as those using template fitting~\cite{2019arXiv191103547S} or solving for additional degrees of freedom that mimic leakages that depend on $\cos\left(2\psi\right)$ and $\sin\left(2\psi\right)$ may help mitigating gain or beam-related biases~\cite{McCallum:2020jsp}, at the cost of an increased noise in the final map.

Mitigating systematic biases on the lensing power spectrum specifically can also be achieved by calculating cross-spectra. Reconstructing the lensing potential using different pairs of maps from different observation runs, frequencies, detector-pair sets, or other data splits, and using these to calculate cross-power spectra, could help in averaging out systematic effects, as each map is affected differently by random systematics and the size of the connected bias terms may be substantially reduced~\cite{Madhavacheril:2020ido}. The resulting power spectrum might be less affected by the systematics, but its uncertainty is likely to increase due to the estimator being less optimal, as well as potentially more issues with missing pixels and other issues affecting map-making using less data.

The lensing reconstruction analysis could also adopt some mitigation techniques. Due to the strong dependency of a lensing reconstruction analysis on the experiment forward modeling through MC simulations, the most direct way of mitigating most biases is by modeling the systematics in the MC simulations used to obtain the debiasing terms, namely MF, $\RDN_{0,L}^{\phi\phi}$ and $\MCN_{1,L}^{\phi\phi}$. In our analysis, the only similarities between the MC and ``data'' simulation sets are the instrument specifications and scanning strategy. Investing resources into more precise modeling of systematics in the MC simulations could reduce their impact on the lensing reconstruction if they are partly simulated. However, not all systematics can be accurately simulated, and some parameter uncertainty in the systematic modeling would not mitigate lensing reconstructions biases entirely. Since the detection level of the systematics-induced lensing biases shown in this work are already low, we do not explore this method of mitigation. We do, however, compare between the MF debiasing terms with and without systematics to understand if any systematics-induced biases may be mitigated by including systematics in the MF simulations.

\begin{figure}[!h]
\centering
\includegraphics[width=\columnwidth]{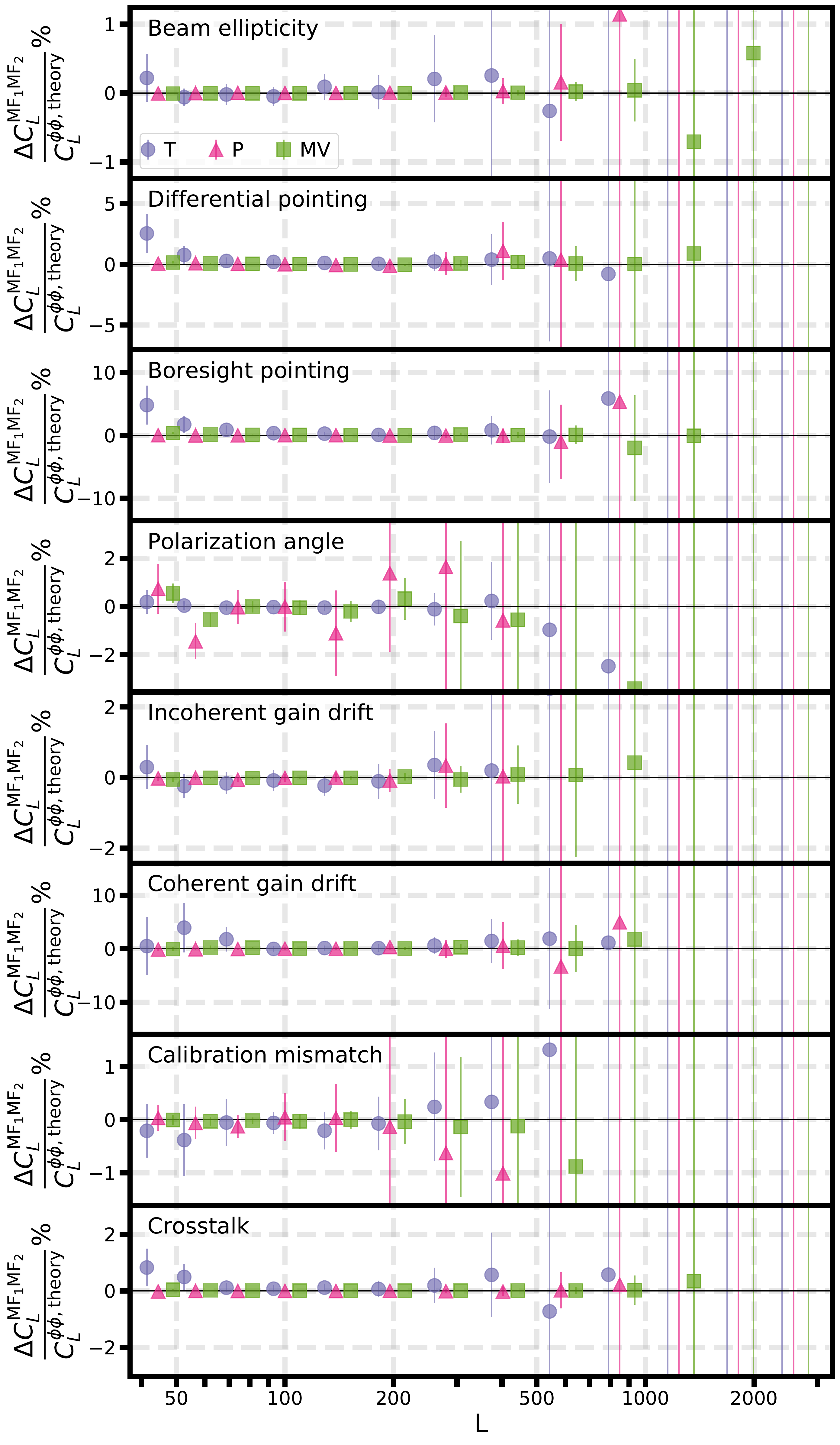}
\caption{The fractional differences between $\Delta C_L^{\MF_1\MF_2}$ and $C_L^{\phi\phi,\theory}$. $\Delta C_L^{\MF_1\MF_2}$ is the difference between the MF cross-spectra calculated using simulations with and without systematics for T (purple circles), P (pink triangles), and MV (green squares) lensing reconstructions.}
\label{fig:MF_comparison}
\end{figure}

Including systematics in the MC simulations will only give a non-zero contribution to the mean field in specific cases, e.g. where the amplitude of the noise mean field is affected by systematics, or where systematics-inducing parameters are known (e.g. the actual beam ellipticities or the crosstalk scheme). Systematics leading to a specific spatial pattern that depends on the specific actual realization or time variation of random variables would average to zero if only random realizations can be simulated. To test whether modeling randomized systematic effects in the MF simulations might help, we use random variables with the same parameter distributions as with the ``data'' sets. The fractional differences between the cross-MF power spectra $C_L^{\MF_1 MF_2}$ with and without systematics and the lensing power spectrum are shown in Fig.~\ref{fig:MF_comparison}. The resulting biases are all consistent with zero, meaning that including a level of variance in the systematics modeling in the MF simulations may not improve the lensing reconstruction. Planck showed that for their specifications (e.g. beam size and scanning strategy) the known beam ellipticity was also negligible when calculating the MF~\cite{Hanson:2010gu} over the multipole range we consider. We have shown that systematic effects from the narrow beams that we considered are relatively negligible for lensing reconstruction, so we do not attempt to model them in the MF simulations. 
The polarization angle and coherent gain drift systematics are more important, but including these systematic effects in the MF simulations with parameter uncertainties similar to those used in the ``data'' simulations did not result in bias mitigation. We did, however, find that including the polarization angle systematic in the MF simulations when reconstructing the curl lensing signal successfully mitigated the signal to undetectable levels.

\begin{figure}[!h]
\centering
\includegraphics[width=\columnwidth]{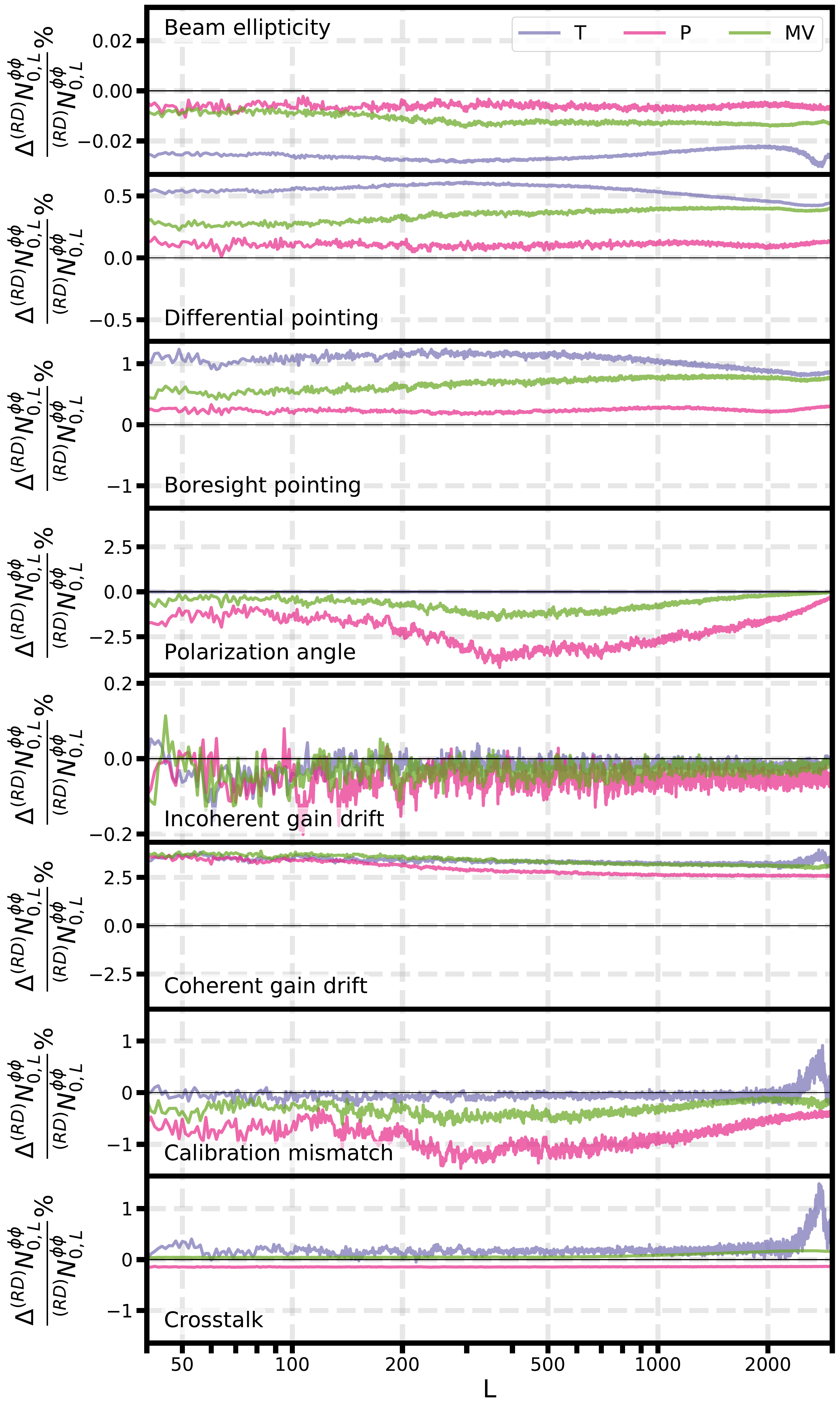}
\caption{The fractional differences between $\RDN_{0,L}^{\phi\phi}$ with and without systematics for T (purple lines), P (pink), and MV (green lines) lensing reconstructions for a given CMB+noise realization ``data'' simulation.}
\label{fig:RDN0_nosyst_vs_RDN0}
\end{figure}

The other debiasing term, $\RDN_{0,L}^{\phi\phi}$, already responds to the ``data'' CMB power spectrum amplitude and shape, and mitigates (to leading order) some of the biases that affect the connected reconstruction noise. The fractional differences between $\RDN_{0,L}^{\phi\phi}$ for a given realization for ``data'' simulations with and without systematics are shown in Fig.~\ref{fig:RDN0_nosyst_vs_RDN0}. The calibration mismatch systematic bias is largely mitigated by the $\RDN_{0,L}^{\phi\phi}$ subtraction, especially for polarization-only reconstruction; its lensing power spectrum-level biases are less than 0.5\% of the lensing power spectrum amplitude while its $\RDN_{0,L}^{\phi\phi}$ amplitude is about 1\% higher than the systematic-free $\RDN_{0,L}^{\phi\phi}$ amplitude. The $\RDN_{0,L}^{\phi\phi}$ amplitude is also affected by the biases of the other systematics, however their lensing spectrum biases remain large compared to the systematics-free spectrum. The largest $\RDN_{0,L}^{\phi\phi}$ amplitude deviation results from the polarization angle and coherent gain drift systematics. While the use of this debiasing term does help with decreasing the lensing spectrum bias, it does not necessarily mitigate it to negligible significance levels.

As an alternative to full modeling of systematic effects in the MC simulations, some biases can be mitigated by using a different fiducial beam in the lensing reconstruction analysis. We have shown in Subsec.~\ref{sec:map_power_biases} that the main beam ellipticity, differential pointing and boresight pointing systematics biases appear as a change in the effective beam model. In practice, the beam model is often determined by dedicated observations of point sources~\cite{Hasselfield:2013zza,Ade:2014afa,Henning:2017nuy}. These empirical measurements include the same observational systematics, so the effective beam model determined from them should already mitigate some of the beam-like effects to some extent.

To test how well an effective beam can mitigate the lensing systematics we, performed lensing reconstructions which included an effective beam. We corrected our baseline Gaussian beam, which has a width $\sigma_{{\scriptstyle \FWHM}}$, by a correction beam with width $\sigma_{\corr}$. The value of $\sigma_{\corr}$ was obtained from fitting the $C_{\ell}^{TT}$ fractions with and without systematics to an effective circular Gaussian beam model. The total width of the effective beam, $\sigma_{\eff}$, is then given by
\begin{equation}
\sigma_{\eff}^{-2} \equiv \sigma_{{\scriptstyle \FWHM}}^{-2} + \sigma_{\corr}^{-2}.
\label{eq:effective_beamwidth}
\end{equation}
More generally, one could define an effective beam transfer function as a general function of $\ell$, but using the simple Gaussian model already captures the main systematic effect.

Boresight pointing jitters are expected to be well captured by an effective beam. We find that the correction determined from the power spectrum gives an effective beam correction $\sigma_{\corr}$ that matches the input pointing distribution width very well (13 arcseconds, matched to about 0.4\%). Applying this effective beam lowers the lensing bias detection levels to 0.06 (T), 0.001 (P), 0.07 (MV). The boresight pointing jitters are not correlated to the scan in the basic model we considered, and do not depend on detector-level properties, which makes this effect simple to mitigate. The main effect should be included automatically in beam measurements from point sources. In more realistic cases, pointing errors might be correlated to the motion of the telescope, or through systematic errors in the pointing solution that relate the recorded position of the telescope encoder to the true sky position. Such correlations could introduce larger biases in lensing reconstruction as they could mimic correlated shifts on the sky, however they are hard to model and quantify in advance as they are highly instrument-dependent. A similar mitigation strategy could, however, reduce the overall effect. An effective beam transfer function known to at least 10-20\% precision should be sufficient to mitigate the boresight pointing bias effectively.

Differential pointing produced an effective-beam like effect, but also $T\rightarrow P$ leakage. This systematic is mainly due to distortions in the focal plane and in the telescope mirror, so the effect is coupled to the scanning strategy and overall cross-linking of different pixels. Employing an effective beam in this case can only partly mitigate the effect, since it would not correct $T\rightarrow P$ leakage. The beam correction is also likely to be less well captured by point-source beam measurements, since point source scans are usually different than the scanning strategy used for CMB observations. If the leakage corrections can be constrained or measured well enough from calibration observations, and they are relatively stable in time, it may be possible to propagate them through simulations to define an effective transfer function that would mitigate most of the effect. We found that by using the effective beam determined from the power spectra, the differential pointing bias detection significance levels decrease to 0.03 (T), 0.02 (P) and 0.01 (MV). These residual detection levels are consistent with the biases expected from having only the leakage terms involving $b_{-}$ of Eq.~\eqref{eq:d_minus_plus_explicit} that give rise to $T\leftrightarrow P$ leakage in the data simulations. Using an effective beam should therefore remove the majority of the differential pointing bias, as the $T\leftrightarrow P$ leakage biases are subdominant. The fitted correction width $\sigma_{\corr}$ deviates from the mean of the differential pointing distribution, 15 arcseconds, by about 12\%.

The beam ellipticity bias mainly originates from the $b_{+}$ leakage terms; removing the $b_{+}$ terms results in biases which are $\sim2$ orders of magnitude smaller, so the majority of this bias is also corrected by an effective beam. Measurements of the beam transfer function from calibration observations should be sufficient to capture the majority of the ellipticity systematic and correct for it. However, there may be some deviations between the ``true'' and measured beam shapes due to the coupling to the scanning strategy.

To assess how accurately the effective beam needs to be known, we used a beam correction width reduced by a factor of 2 from the best value to correct for the differential pointing bias. This lead to a reduction of the bias detection levels by about a factor of 2 compared to using no effective beam. For boresight pointing, the effective beam is likely to be measured better than this, as calibration observations are expected to estimate the correct beam shape for this systematic quite well; however, our results suggest that even an approximate beam model may be sufficient to substantially reduce the lensing biases.

The most common method for mitigating polarization angle systematics is by fitting the resulting non-zero $EB$ cross-spectrum to the analytical expression~\cite{2013ApJ...762L..23K},
\begin{equation}
\begin{split}
C_\ell^{\tilde{E}\tilde{B}} &= \frac{1}{2}\sin\left(4\Delta\psi\right)\left[C_\ell^{EE}-C_\ell^{BB}\right],
\end{split}
\label{eq:pol_ang_powers}
\end{equation}
where $\tilde{X}$ is a polarization field affected by the polarization angle systematic for a constant angle shift $\Delta\psi$. To test the effectiveness of this mitigation method, we fitted the resulting $EB$ power spectrum, shown in Fig.~\ref{fig:polang_eb_curl}, to the analytic formula of Eq.~\eqref{eq:pol_ang_powers} using the theoretical $E$ and $B$ power spectra to obtain the effective angle $\Delta\psi$, and used it to rotate the input $Q$ and $U$ maps for the lensing analysis. Although this method approximates the systematic to be a global map-level effect, while the systematic is in practice injected at the per-detector level, it corrects for most of the effect and considerably mitigates both the lensing biases and the curl signal to undetectable levels. After mitigation, the lensing bias detection levels reduce to $0.04\sigma$ (P) and $0.03\sigma$ (MV). This mitigation does not affect the already-negligible bias detection levels for a temperature-only reconstruction significantly. We note that this mitigation strategy makes assumptions on the underlying cosmology, i.e. $C_\ell^{\tilde{E}\tilde{B}}=0$ in absence of systematics. This suggests that it could also unwantedly remove any signals which are caused from other sources, such as cosmic birefringence. 
As this systematic produces a lensing curl signal, which could potentially also be coupled with cosmic birefringence, it may be possible to include the curl signal as an additional diagnostic tool to break the degeneracy between systematic-induced and cosmologically-induced rotations.

Gain systematic effects can also be mitigated to some extent, following the common practice of most CMB ground-based experiments, by cross-correlating the resulting maps with external data sets such as the Planck maps~\cite{2011ApJ...740...86H,Louis:2014aua,Ade:2017uvt,Choi:2020ccd}. This calibration can help correct an overall mean gain error, however position-dependent gain variations may still remain after this absolute gain calibration.

In this work, we have we used QEs to assess the effect of instrumental systematics on the lensing reconstruction. Other reconstruction methods could have different sensitivities to these effects, and some may be able to mitigate systematic effects, at least partially. For example, one could in theory produce a bias-hardened reconstruction which is less sensitive to various systematic effects by construction~\cite{2013MNRAS.431..609N}. Methods which avoid the need for reconstruction bias subtraction~\cite{2010arXiv1011.4510S} may also be less sensitive to instrumental systematics, as they do not need to accurately model systematics that affect debiasing terms that are no longer needed. Performing a lensing reconstruction from split data may also prove to be useful against systematics biases, as different data points are affected differently by systematics (and uncorrelated instrument noise) such that their effect on the lensing contractions may average out~\cite{Madhavacheril:2020ido}. Nonetheless, the lensing pipeline we used proved to perform well against these biases, as most of their effects could be corrected directly by $\RDN_{0}$ subtraction, by implementing an effective beam in the lensing analysis, or by rotating the polarization maps prior to the lensing analysis.

\section{Conclusions and future prospects}
\label{sec:conclusions_and_future_prospects}

In this work, we explored how various instrumental systematics affect the lensing reconstruction power spectrum. We reconstructed the lensing potential from CMB simulations that include realistic levels of contamination due to different instrumental systematics expected for an SO-like instrument, and assessed the significance of the resulting biases. We showed that for the instrument specifications and scanning strategy we used, most of the systematics we considered will have a relatively small effect on lensing reconstruction for upcoming CMB experiments, with significance levels of up to $0.5\sigma$, apart from the boresight pointing, polarization angle and coherent gain drift systematics, which produce biases with $>0.5\sigma$ significance levels when left unmitigated. We also investigated whether these instrumental systematics produce a lensing curl signal, and found that when not calibrating for polarization angle errors the signal can be detectable. All of the significance levels we have presented in this work might be somewhat different for a full observation run, especially for gain drifts and calibration mismatch where the biases average out with time. Systematics that appear in the maps' power spectra as an effective beam are also likely to be substantially mitigated once the beam is empirically calibrated.

Future CMB experiments, such as CMB-S4, which will produce CMB maps with even lower instrument noise, may be more sensitive to these systematics, as their lensing reconstruction noise level is expected to be even lower. A more accurate quantitative assessment of the impact of instrumental systematics on lensing reconstruction for a given experiment depends on the details of its scanning strategy, focal plane configuration and instrument properties. These can only be characterized in their full complexity during the observational campaign. As such, the absolute value of the systematics we explored may differ from the results presented in this work. Nonetheless, our results are a useful guide toward identifying the most relevant potential problems and planning the lensing analysis for upcoming ground-based CMB experiments.

Throughout this work we made various simplifications and assumptions, which should be investigated more carefully in the future. Modeling multi-frequency bolometers would allow for a more robust estimator, with some handle on frequency-dependent systematics and foregrounds. However, having more than one band potentially increases the range of possible systematics, and some small systematics could become relatively more important due to the process of foreground cleaning. We also neglected any foreground residuals in the post-cleaning CMB maps and any systematics that may couple to bright foreground emission.
Another effect which may be crucial for upcoming ground-based CMB experiments is correlated noise. When observing the sky from the ground, the resulting time streams are contaminated by atmospheric emission. This could have a significant impact on the resulting CMB maps, which could then bias the lensing reconstruction. Residuals from various correlated noise cleaning methods could also negatively affect the lensing reconstruction. Additional filtering on the map level could improve the reconstruction accuracy from correlated-noise-contaminated maps~\cite{Sherwin:2016tyf}, however a full optimization analysis for such methods has not yet been performed. Exploring systemically how filtering affects lensing biases specifically would require implementing a more sophisticated map-making method than the one we used here~\cite{deGasperis:2016fgd}, and simulation tests would require realistic atmospheric noise simulations~\cite{Errard:2015twg}.

Apart from modeling different systematics, testing how the biases change for different scanning strategies could also be useful for planning optimal scans for future experiments. Many instrumental systematics effects are mitigated when the same area of the sky is scanned multiple times from different angles during several observation runs. Since scanning more area of the sky is also beneficial to reduce cosmic variance, performing an optimization analysis to understand how this interplay affects systematics could be key for future CMB experiment planning.

Another important aspect of CMB lensing is the ability to delens CMB maps with high precision. Systematics may affect a delensing analysis somewhat differently, and some may even prove to be relatively more important for delensing rather than for the lensing power spectrum. Performing a delensing analysis using CMB maps which include systematics would be the next step toward a comprehensive investigation on the effects of systematics on lensing-related analyses.

So far, our results show a promising future for lensing-related CMB cosmology. We have demonstrated that most of the systematics we considered should be relatively negligible for an SO-like experiment, especially when using many more detectors compared to our analysis and observing more sky area and for longer times, or could be mitigated effectively. Within the limitations of our work, we conclude that the upcoming generation of instruments such as SO should be able to deliver the lensing science case they target.

\section{Acknowledgments}
We thank the referee for providing useful comments and suggestions for improving our manuscript. We thank Clara Verg\`{e}s and Max Silva-Feaver for helping with modeling the crosstalk systematic and for useful discussions, Kevin Crowley and Matthew Hasselfield for helpful insights and suggestions regarding gain drifts systematics, Michele Limon and Grant Teply for useful guidelines for modeling the boresight pointing systematic. We thank Sara Simon for guidance on SO systematics and for very useful comments on the manuscript, and the SO calibration, sensitivity and systematics technical working group for suggesting us parameter estimates for the various systematics considered in this work. We acknowledge support from the European Research Council under the European Union's Seventh Framework Programme (FP/2007-2013) / ERC Grant Agreement No. [616170], and support by the UK STFC grants ST/P000525/1 (AL) and ST/T000473/1 (AL and GF). GF also acknowledges the support of the European Research Council under the Marie Sk\l{}odowska Curie actions through the Individual Global Fellowship No.~892401 PiCOGAMBAS.
This research used resources of the National Energy Research Scientific Computing Center (NERSC), a U.S. Department of Energy Office of Science User Facility operated under Contract No. DE-AC02-05CH11231. Some of the results in this paper have been derived using the healpy/HEALPix package~\cite{Gorski:2004by}, and NumPy~\cite{Zonca2019,harris2020array}, SciPy~\cite{2020SciPy-NMeth} and Matplotlib~\cite{4160265} libraries.

\bibliography{systematicsbib,texbase/antony}
\end{document}

%% file: macros.tex
\newcommand{\mksym}[1]{\ifmmode {\rm #1}\else #1\fi}

\providecommand{\text}[1]{\rm{#1}}

\providecommand{\CAMB}{{\tt camb}}

\newcommand{\begm}{\begin{pmatrix}}
\newcommand{\enm}{\end{pmatrix}}

\newcommand\ba{\begin{eqnarray}}
\newcommand\ea{\end{eqnarray}}
\newcommand\bea{\begin{eqnarray}}
\newcommand\eea{\end{eqnarray}}

\newcommand\be{\begin{equation}}
\newcommand\ee{\end{equation}}

\newcommand{\mA}{\bm{A}}

\newcommand{\mC}{\bm{C}}

\newcommand{\mL}{\bm{L}}

\newcommand{\mN}{\bm{N}}

\newcommand{\boldvec}[1]{{\mbox{\boldmath{$#1$}}}}

\newcommand{\vd}{\boldvec{d}}

\newcommand{\vn}{\boldvec{n}}

\newcommand{\vs}{\boldvec{s}}